\documentclass[aps,pre,twocolumn,superscriptaddress,showkeys]{revtex4-2}

\pdfoutput=1

\usepackage[T1]{fontenc}
\usepackage{amsmath,amssymb}
\usepackage{graphicx}
\usepackage{dcolumn}
\usepackage{bm}
\usepackage{float}
\usepackage{color}
\usepackage{amsmath}
\definecolor{rojo}{rgb}{1,0,0}
\definecolor{verde}{rgb}{0,0.8,0.2}
\definecolor{azul}{rgb}{0,0,1}
\definecolor{rosa}{cmyk}{0,1,0,0}
\usepackage{appendix}
\usepackage{amsmath}
\usepackage{amsthm}
\usepackage{amsfonts}
\usepackage{amssymb}
\usepackage{color}
\usepackage{epsfig}
\usepackage{hyperref}
\hypersetup{colorlinks=true, citecolor=blue}
\usepackage{soul}
\usepackage{comment}

\newcommand{\jp}[1]{{\color{red}#1}}

\begin{document}
\preprint{APS/123-QED}

\title{Percolation of a rod-like particle in a static bed of spheres: trapping and passing}

\author{Juan C. Petit}
\email{corresponding author: jcpetit71@gmail.com}
\affiliation{Department of Mechanical Engineering, Northwestern University, Evanston, IL 60208, USA}

\author{Julio M. Ottino}
\affiliation{Department of Mechanical Engineering, Northwestern University, Evanston, IL 60208, USA}
\affiliation{Department of Chemical and Biological Engineering, Northwestern University, Evanston, IL 60208, USA}
\affiliation{Northwestern Institute on Complex Systems (NICO), Northwestern University, Evanston, IL 60208, USA}

\author{Richard M. Lueptow} 
\affiliation{Department of Mechanical Engineering, Northwestern University, Evanston, IL 60208, USA} 
\affiliation{Department of Chemical and Biological Engineering, Northwestern University, Evanston, IL 60208, USA}
\affiliation{Northwestern Institute on Complex Systems (NICO), Northwestern University, Evanston, IL 60208, USA}

\author{Paul B. Umbanhowar}

\affiliation{Department of Mechanical Engineering, Northwestern University, Evanston, IL 60208, USA}

\date{\today}
\begin{abstract}
We numerically investigate percolation of independent frictionless glued-sphere rod-like particles under gravity through a disordered static bed of larger spheres. We identify two distinct regimes: a \emph{trapping} regime, where rods stop after percolating a limited distance in the bed and a \emph{passing} regime, where rods percolate continuously with constant mean velocity. The transition between these regimes is governed by the length of the rod and the geometrical trapping threshold for spherical particles based on the rod diameter and the minimum pore throat diameter defined by three touching large spheres. The percolation velocity for all rod geometries, including the single sphere limit, collapses onto a single curve when scaled with the gravitational acceleration and the bed sphere diameter. The results also demonstrate that short rods percolate nearly twice as fast as long rods due to the geometric constraints associated with the disordered pore structure of the static bed. Consequently, long rods are more susceptible to trapping via specific contact configurations with the bed spheres, which differ from those for short rods. These results reveal how shape anisotropy introduces dynamical constraints and thresholds in granular percolation, with implications for predicting segregation in mixtures of non-spherical particles.
\end{abstract}

\maketitle

\section{Introduction}

The percolation of fine particles, or fines, through a bed of static larger particles is relevant in many scientific, geophysical, and engineering contexts. Fine particle percolation is a subset of the larger phenomenon of segregation, which refers to the spontaneous separation of particles with different physical properties—such as size, density, shape, or surface roughness—typically when subjected to flow, vibration, or other forms of mechanical agitation \cite{ottino2000mixing, umbanhowar2019modeling, jones2021Predicting, Pohlman2006Surface, tripathi2013density, kudrolli2004size, jaeger1996Granular, duran2012sands}. However, segregation can also occur in the absence of agitation when fine particles are small enough to percolate through the interstices between static large particles due to gravity alone, a phenomenon known as \emph{free sifting}. Segregation is ubiquitous yet difficult to predict due to its inherently non-equilibrium and highly-nonlinear nature. For instance, in geophysical contexts, segregation shapes the structure of landslides, debris flows, and avalanches, where coarse material rises to the surface while finer material settles or migrates downward, influencing mobility and run-out distance \cite{johnson2012grain, marks2015mixture, phillips2006enhanced}. In industry, segregation is a persistent challenge wherever powders or grains are handled: in pharmaceuticals, it leads to dosage variability when fine active ingredients separate from excipients \cite{jakubowska2021blend, dave2015excipient, sarkar2017role}; in food and agriculture, uneven distribution of small grains or additives affects product quality and appearance \cite{jian2019segregation, flore2009aspects}. Thus, segregation is simultaneously a natural sorting mechanism and a technological hurdle, making its study crucial for both understanding geological processes and optimizing industrial operations.

Segregation related to free sifting is common in industrial solids handling processes. It comes about in situations where large particle size differences occur, for instance, as a consequence of the inherent polydispersity of the granular material, particle breakage, or the intentional addition of particles such as free-flow agents. However, the presence of fines in granular systems can have undesirable outcomes such as degraded product quality, equipment fouling, health risks from inhalation, and safety hazards such as dust explosions \cite{bemrose1987review, schulze2008powders}.

Although we consider the percolation of rod-like particles through a static bed of spherical particles in this paper, it is helpful to first provide background on the percolation of fine spherical particles. The critical size ratio above which a small sphere can pass through the smallest interstices in a bed of large spheres is based on a triangular configuration of three contacting large spheres with diameter $d_{\mathrm{L}}$. The largest sphere that fits inside the pore-throat of this configuration has diameter $d_{\mathrm{t}} = d_{\mathrm{L}} / R_{t}$, where $R_{t} = \sqrt{3}/(2-\sqrt{3}) = 6.464$, which is known as the geometrical trapping size ratio, see Fig.~\ref{fig1}(a). Thus, small spherical particles of diameter $d_{\mathrm{S}}$, freely percolate through a static bed of large particles for $R = d_{\mathrm{L}}/d_{\mathrm{S}}$ > $R_t$ \cite{dodds1980porosity, lomine2010transit, rahman2008simulation}. For $R < R_{t}$, small spheres may percolate for some distance in a random static packing of larger spheres, but they eventually become trapped. This value of $R_t=6.464$ applies to monodisperse large particle beds; any polydispersity or deformability of the large particles decreases $R_t$~\cite{vyas2024impacts}. Beyond simple percolation, this geometric threshold serves as the physical origin for the decoupling of jamming transitions in dense mixtures, acting as the critical limit that allows the large-particle backbone to arrest independently, while an additional jamming transition of the fine species emerges upon further compression \cite{petit2020additional, petit2023structural}.

Several recent studies have investigated the percolation dynamics of fine spheres in both static and sheared beds of monodisperse large spheres above the trapping threshold $R_{t}$ \cite{gao2024vertical, gao2023percolation, vyas2026fine}. Following the nomenclature of these studies, particles are defined as \emph{fine} particles for $d_{\mathrm{S}} < d_t$, and \emph{small} particles for $d_t < d_S < d_L$.  In static beds, the percolation velocity decreases as the packing fraction increases, since denser packing reduces pore throat sizes and thus limits percolation. However, the percolation velocity increases with increasing $R$ because smaller fine particles more easily pass through the pores \cite{gao2023percolation}. The percolation velocity of fine spheres is strongly influenced by the coefficients of restitution, $e,$ and friction, $\mu$~\cite{vyas2026fine}. For frictionless particles, the percolation velocity decreases with increasing $e$ because fine particles collide with bed particles more frequently due to a corresponding increase in their fluctuation velocity, slowing their downward progress. With increased $\mu$, the percolation velocity is reduced at low $e$ because of increased frictional losses, but enhanced at high $e$ due to reduced excitation of the fine particles. In sheared beds of large particles, the percolation velocity of small spheres exhibits a bell-shaped dependence on the shear rate, reaching a maximum at a specific shear rate that depends on $R$~\cite{gao2024vertical}. Below the maximum, increasing shear opens gaps between large particles to allow small particles to percolate through the bed, while above the maximum, increasing shear increases small particle velocity fluctuations, thereby slowing their downward percolation. 

Yet real-world granular materials and natural sediments rarely consist of perfect spheres. Rather, they often contain rods, fibers, or flakes—particles whose elongated shapes introduces orientation, interlocking, and different types of contacts that are absent in spherical systems. The distinction between spheres and rods matters. A rod can only pass through a void if it is aligned with respect to the void. As a result, rod percolation is strongly coupled to geometric constraints. By studying how frictionless rods percolate through beds of large spheres, we consider how geometric constraints affect percolation in order to shed light on how particle anisotropy influences percolation.

This paper is organized as follows. In Sec.~\ref{SecII}, we describe the numerical protocol as well as how the packed bed of spherical particles and the rod particles are generated. Section~\ref{SecIII} includes results for both the ``trapping'' regime, where rods become stuck in the static bed after percolating some distance, and the ``passing'' regime, where frictionless rods percolate freely through the static bed of spherical particles. The paper concludes with a summary of the main findings.

\section{Numerical simulations}
\label{SecII}

Discrete element method (DEM) simulations are carried out using MercuryDPM \cite{weinhart2020fast} to study the percolation of fine rod-like glued-sphere particles through a disordered bed of large spheres. We use the linear spring-dashpot model to characterize normal forces: ${\it \bf f}^n_{ij}=f^n_{ij} \hat{\bf n} = ( k_{n} \alpha_n^{ij} + \gamma_{n} \dot\alpha_n^{ij}) \hat{\bf n}$ \cite{cundall1979discrete, luding2008cohesive, petit2022bulk}, where $\alpha_n^{ij}$ and $\dot\alpha_n^{ij}$ are the contact overlap and relative velocity in the normal direction $\hat{\bf n}$ between particles $i$ and $j$, $k_{n}$ is the contact spring stiffness, and $\gamma_{n}$ is the damping coefficient. Tangential contact forces are set to zero to focus on geometric effects. The rods and bed spheres share the same material properties: $\rho = 1\,\mathrm{g/cm^{3}}$, $k_{n} = 10^{5}\,\mathrm{dyn/cm}$, and $\gamma_{n} = 0.47 \,\mathrm{g/s}$.

\subsection{Packed bed preparation}

\begin{figure}[t]
\centering \includegraphics[scale=0.32]{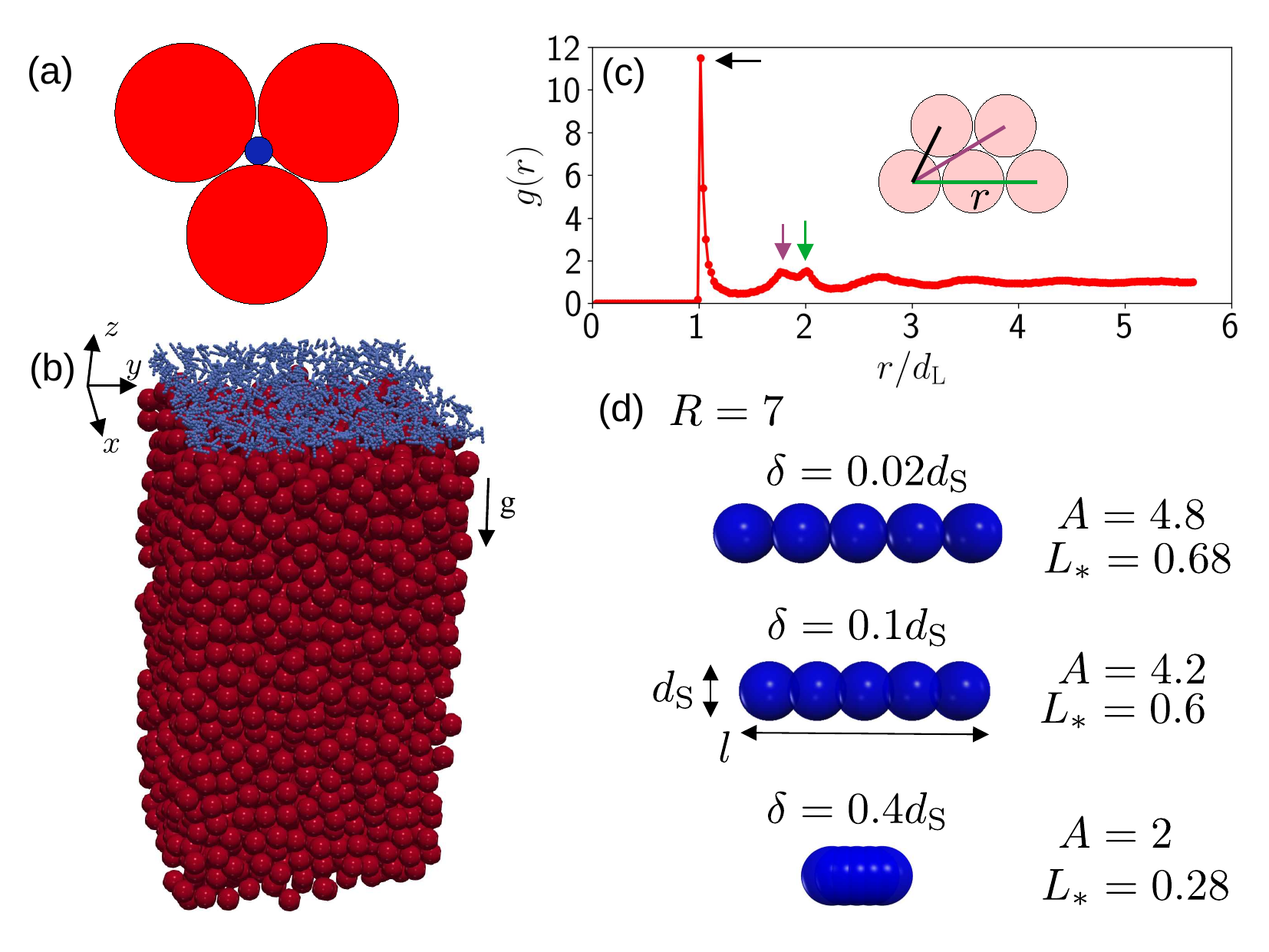}
\caption{(a) Triangular configuration of three contacting large spheres with diameter $d_\mathrm{L}$ (red) defining the pore throat diameter $d_t$ (blue) associated with the geometric trapping limit, $R_t$. (b) Randomly packed bed of large spheres ($\phi = 0.60$) with initial configuration of non-interacting rods (blue) at the top. (c) Radial distribution function, $g(r)$, of the packed bed showing a disordered configuration beyond nearby particles for $r > 2 d_\mathrm{L}$. (d) Examples of rods with various aspect ratios, $A = d_{\mathrm{L}}/d_{\mathrm{S}}$, and length ratios, $L_{*} = l/d_{\mathrm{L}}$.}
\label{fig1}
\end{figure}

The static packed bed is formed by randomly placing $3000$ frictionless non-overlapping spheres of diameter $d_{\mathrm{L}} = 2 \,\mathrm{mm}$ with zero velocity in a box with initial dimensions $20d_{\mathrm{L}}\times20d_{\mathrm{L}}\times36d_{\mathrm{L}}$ and initial volumetric packing fraction $\phi = 0.1$. To avoid wall effects, periodic boundary conditions are applied in all directions. The spheres are then compressed under zero gravity to a maximum packing fraction of $\phi_{\mathrm{max}} = 0.65$ and then decompressed until $\phi = 0.60$. The compression and decompression phases are executed via continuous isotropic scaling of the simulation boundaries at each time step \cite{petit2022bulk, luding2022understanding, santos2024protocol, o2003jamming}. During compression, the box dimensions are reduced at a constant strain increment of $\epsilon=10^{-3}$ per time step, driving the system toward $\phi_{\mathrm{max}} = 0.65$ at $t \approx 0.63 \sqrt{d_{\mathrm{L}}/g}$. Once $\phi_{\mathrm{max}}$ is reached, the system is allowed to relax. After relaxation, the decompression phase is conducted five times slower ($\epsilon= 2 \times 10^{-4}$) to reach the final target of $\phi = 0.60$ over a total time of $t \approx 12 \sqrt{d_{\mathrm{L}}/g}$. Throughout both compression and decompression, a background dissipation is applied to particle velocities to accelerate the removal of kinetic energy. The relaxation is sufficiently long for the kinetic-to-potential energy ratio of the bed spheres to fall below $10^{-4}$. Then, the positions of the spheres are fixed to prevent any motion caused by interactions with the rods percolating through the static bed. Figure~\ref{fig1}(b) shows the packed bed of large spheres formed after decompression with $\phi = 0.60$ and final dimensions of $11.33 d_{\mathrm{L}} \times 11.33 d_{\mathrm{L}} \times 20.39 d_{\mathrm{L}}$. Structural analysis via the radial distribution function [(Fig.~\ref{fig1}(c)] verifies the bed's disordered nature. A nearest neighbor peak occurs at $r/d_{\mathrm{L}} = 1$ (black arrow), while much smaller peaks at $r/d_{\mathrm{L}} \approx 1.73$ (purple arrow) and $r/d_{\mathrm{L}} = 2$ (green arrow) correspond to local hexagonal configurations, as indicated in the inset. At larger distances, $g(r) \to 1$, indicating a disordered configuration.

\subsection{Rod-like geometry}

A rod consists of $n$ collinear small spheres with diameter, $d_{\mathrm{S}} = d_{\mathrm{L}}/R$, with a geometry dependent overlap $\delta$ \cite{ostanin2024rigid, parteli2013simulation}. The maximum length of a rod, measured from the ends of the rod at $\delta = 0$, is $l_{\rm{max}} = nd_{\mathrm{S}}$. Different rod lengths, $l < l_{\rm{max}}$, with aspect ratio, $A = l / d_{\mathrm{S}}$, are obtained by increasing the overlap $\delta$ between small spheres for a fixed $n$: $l(\delta) = n d_{\mathrm{S}} - (n - 1)\delta$. The rod density is set equal to the bed particle density, $\rho$, with the rod mass modeled as a uniform cylinder: $M = \pi d^{2}_{\mathrm{S}}l\rho/4$. To efficiently sample the parameter space while minimizing computational overhead, we model rods with $1.5 \leq A < 5$ utilizing five constituent spheres, and rods with $A \geq 5$ utilizing eleven. Figure~\ref{fig1}(d) illustrates representative geometries for rods with $n=5$ and $R = 7$ for three different overlaps: $\delta/ d_{\mathrm{S}} \in \{0.02, 0.1, 0.4\}$. We explore aspect ratios $1 \leq A \leq 10$ for each size ratio $R \in \{5, 6, 6.46, 7, 8, 9, 10\}$. The dimensionless length scale, $L_{*} = A / R = l / d_{\mathrm{L}}$, is varied over the range $0.1 \leq L_{*} \leq 1$. It reflects the geometric constraint of the rod length relative to the bed pore structure (characterized by $d_\mathrm{L}$), while $R$ captures how a rod fits through a pore throat based on the rod diameter.

To investigate the percolation of rod-like particles, we simulate systems in which $N = 1000$ non-interacting frictionless rods are dropped into a static bed of frictionless large spheres, the initial condition for which is illustrated in Fig.~\ref{fig1}(b). Simulations with $N = 5000$ rods are used in selected cases to reduce statistical uncertainty. Rod-rod contacts are disabled, whereas rod-sphere interactions are retained. In this way, rods do not interact with each other and follow a path based only on interactions with large spheres. Consequently each rod trajectory can be seen as an independent simulation. The rods are initially positioned at a distance $d_{\mathrm{L}}$ above the surface of the bed with random orientations and positions distributed across the entire transverse area of the simulation box, see Fig.~\ref{fig1}(b). The periodic boundary at the top of the computational domain is initially removed to let the rods fall into the static bed under gravity and is then restored after all rods have entered the bed so that they may re-enter the top of the simulation domain multiple times.

During percolation, rods interact with bed particles through their constituent small spheres, governed by the linear spring-dashpot contact model described above. While the normal damping coefficient, $\gamma_n$, remains constant, the restitution coefficient, $e$, is computed internally by MercuryDPM at the level of individual sphere-sphere contacts via $e = \exp(-\gamma_{n} t_{c} / 2 m_{\mathrm{eff}})$, where $t_c = 2 \pi m_{\rm eff} / \sqrt{4 m_{\rm eff} k_n - \gamma_n^2}$ represents the characteristic contact duration. The simulation time step is then chosen as $t_{c}/50$. The dynamics are governed by the effective mass of the contacting pair, $m_{\mathrm{eff}} = m_{\mathrm{L}} m_{\mathrm{S}}/ (m_{\mathrm{L}} + m_{\mathrm{S}})$, where $m_{\mathrm{L}}$ and $m_{\mathrm{S}}$ are the masses of the contacting large bed particle and constituent small particle of the rod, respectively. For the parameters studied here, the resulting restitution coefficient for sphere-sphere contacts is $e \leq 0.6$. This bound ensures sufficiently inelastic collisions, allowing rods to dissipate kinetic energy rapidly rather than developing an excessive granular temperature that can reduce the percolation velocity~\cite{vyas2026fine}.

Similar to the methodology used in previous studies of rod-like particles \cite{kodam2009force}, the effective restitution of the multi-sphere rod assembly, $e_{\mathrm{rod}}$, is approximated as the individual sphere-sphere coefficient, $e,$ for single contact based on $\gamma_n$, which is the same for bed particles and constituent rod particles. This simplification is justified because the energy dissipation of the rod is captured through the interaction between a small sphere that is part of the rod and a bed large sphere. While $e$ is calculated based on these local sphere-sphere interactions, the effective restitution of a non-spherical particle is typically lower than that of a single sphere \cite{vyas2023modelling, kodam2009force, glielmo2014coefficient} because energy dissipation depends strongly on impact orientation, shape, and size—factors that activate rotational modes and potential multi-point contacts during a collision.

\section{Results}

\label{SecIII}

\subsection{Trapping regime}
\label{SecIIIA}
As a rod falls through the static bed it can become trapped if it is sufficiently long or if its constituent spheres are sufficiently large. In this section we study the trapping behavior of rods by characterizing the depth to which a rod penetrates the bed before becoming trapped, the number of contacts it has when trapped, the location of those contacts on the bed particles, the orientation of trapped rods, and details of the transition between continuous percolation and trapping.

\subsubsection{Characteristic penetration length}
\label{SecIIIA1}

A randomly oriented rod released from an arbitrary horizontal position at a fixed height above a bed of large spheres [e.g., as shown in Fig.~\ref{fig1}(b)] will fall into the bed and percolate continuously or become trapped after falling a finite distance depending on the rod and bed geometries. For geometries where trapping occurs, the trapping process is probabilistic as it depends sensitively on the rod's initial conditions---identical rods released at different horizontal locations and different orientations penetrate the bed to different degrees. 

\begin{figure}[t]
\centering
\centering \includegraphics[scale=0.5]{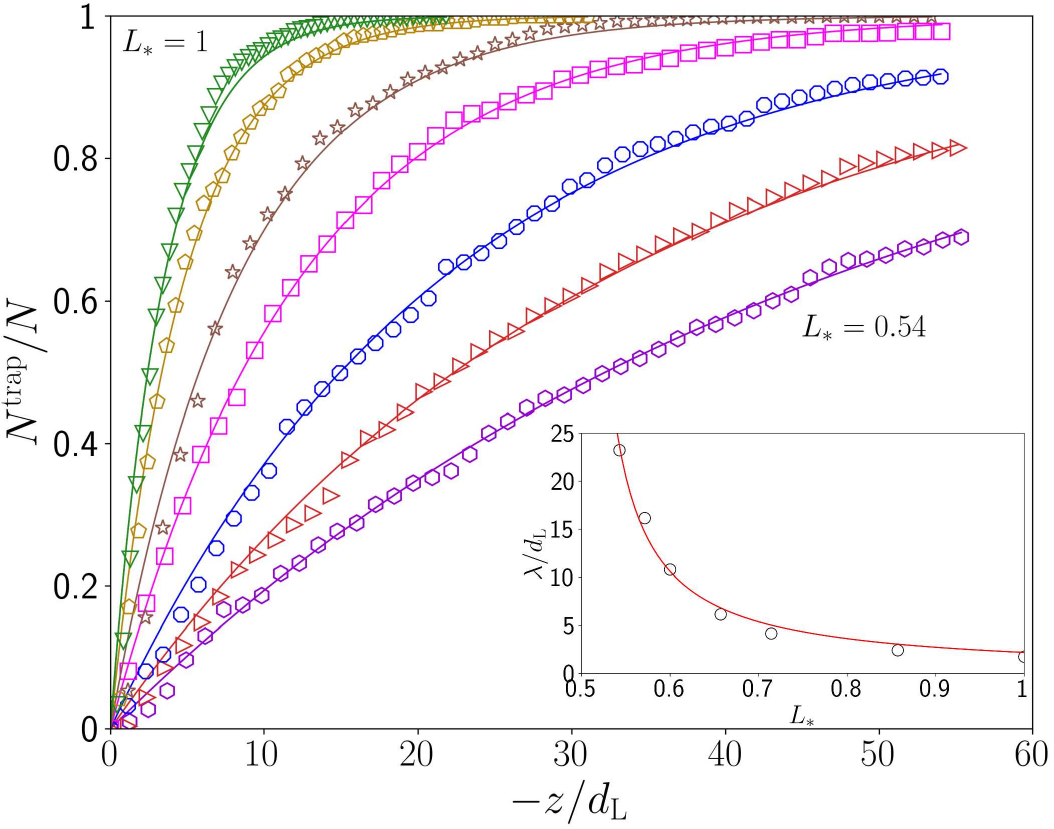}
\caption{Fraction of trapped rods, $N^{\mathrm{trap}}/N$, vs scaled depth, $z/d_{\mathrm{L}}$, for $R = 7$ and $L_{*} \in \{0.54, 0.57, 0.6, 0.65, 0.71, 0.86, 1\}$. Solid curves are fits to $N^{\mathrm{trap}}/N = 1 - \exp(z/\lambda)$. Inset: Scaled characteristic bed penetration length, $\lambda/d_\mathrm{L}$, vs scaled rod length, $L_{*}$; solid curve is a fit to $\lambda/d_\mathrm{L} = c/(L_{*} - L^c_{*})$, where $L^c_{*} \approx 0.52$ is the critical trapping length and $c \approx 1.1$.}
\label{Numtrap_lstar_meanz}
\end{figure}

To characterize the trapping behavior, we determine the trapping probability as a function of depth. A rod is considered trapped when the magnitude of its vertical velocity $\left|v_{z}\right|< 2.2 \times 10^{-4} \sqrt{g d_{\mathrm{L}}} \approx 3 \times 10^{-3}$\,cm/s (results are relatively insensitive to the specific value of this velocity threshold). Rod trajectories that pass through the bed periodically are unwrapped. The trapping probability for an ensemble of $N$ rods is given by $N^{\mathrm{trap}}/N$, where $N^{\mathrm{trap}}(z)$ is the total number of rods trapped above depth $z.$ Figure~\ref{Numtrap_lstar_meanz} plots $N^{\mathrm{trap}}/N$ as a function of scaled depth, $-z/d_\mathrm{L},$ for $R=7$. For the longest rod studied with $L_*=1$, the trapping probability is nearly one above a depth of $\approx 15 d_\mathrm{L}$. As the rod length is decreased from $L_{*}=1$, the depth at which all rods are immobilized ($N^{\mathrm{trap}}/N=1$) increases. Rods with $L_{*} \leq 0.43$ are never trapped.

The probabilistic nature of the trapping process and the data in Fig.~\ref{Numtrap_lstar_meanz} suggest fitting the data to an exponential form: 
\begin{equation}
 N^{\mathrm{trap}}/N = 1 - \exp(z/\lambda),
 \label{eq:expfit}
\end{equation}
where $\lambda$ represents the characteristic penetration length at which $63\%$ of the rods are trapped or, equivalently, where a single rod has a $63\%$ chance of being trapped. This expression provides a good fit to all of the data shown, is effectively independent of the number of constituent spheres forming the rod, and allows us to characterize the dependence of $\lambda/d_\mathrm{L}$ on $L_*$ as shown in the figure inset. As $L_*$ is increased, $\lambda/d_\mathrm{L}$ approaches zero as would be expected since very long rods cannot penetrate the bed. More interestingly, as $L_*$ is decreased $\lambda/d_\mathrm{L}$ appears to diverge at a characteristic rod length of $L_*^c$ marking the transition between trapping and continuous percolation. Fitting the dependence of $\lambda/d_\mathrm{L}$ on $L_*$ to 
\begin{equation}
 \lambda/d_{\mathrm{L}} = c/(L_{*} - L^c_*)
 \label{eq:criticalL}
\end{equation}
accurately captures the behavior (solid red line in inset of Fig.~\ref{Numtrap_lstar_meanz}) and provides a value of $L_*^c\approx 0.52$ for this case with $R=7$. 

\begin{figure}[t]
\centering \includegraphics[scale=0.52]{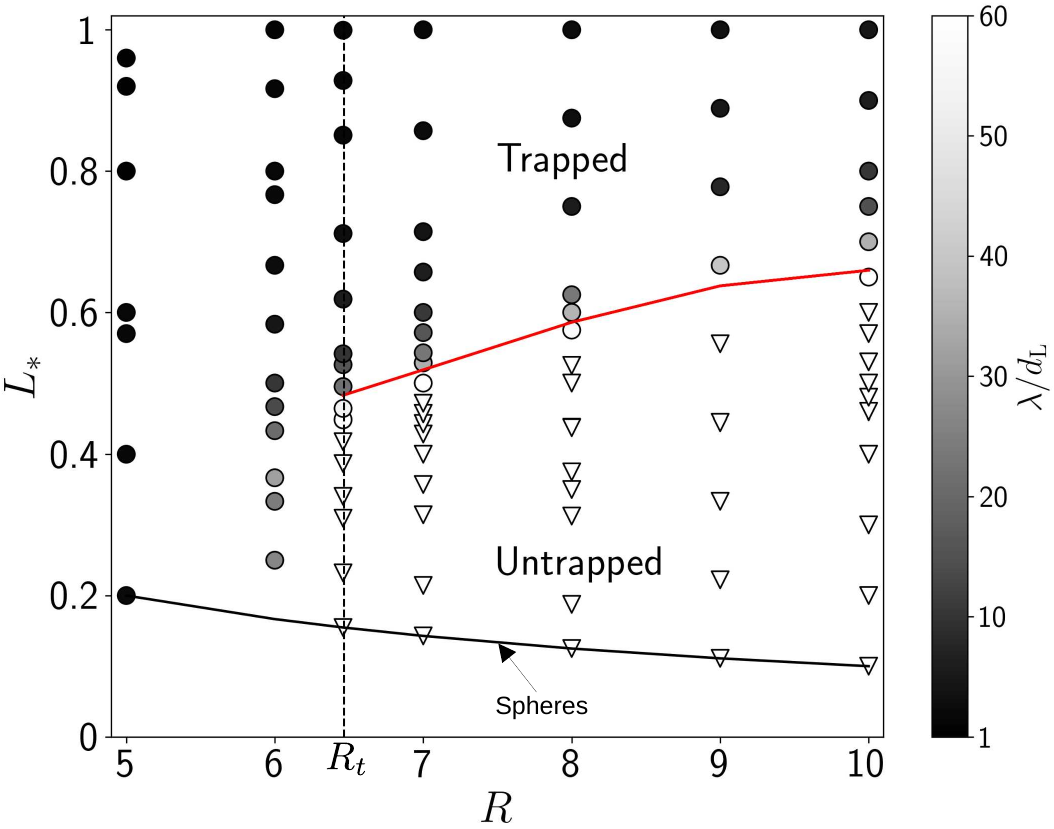}
\caption{Phase diagram of trapped and passing regimes vs rod length, $L_{*},$ and size ratio, $R$, in a disordered bed of spherical particles with $\phi = 0.60$. Downward triangles represent the continuous percolation regime while circles indicate the trapping regime, characterized by $\lambda/d_{\mathrm{L}}$ (grayscale), which diverges at the boundary between trapped and untrapped rods. Solid black and red lines denote, respectfully, the spherical limit ($L_{*} = R^{-1}$) and the critical trapping length, $L_{*}^{c}$, extracted from fits to Eq.~\ref{eq:criticalL} (see inset in Fig.~\ref{Numtrap_lstar_meanz}). The vertical dashed line indicates the minimum pore-throat-based geometric trapping threshold at $R = R_{t} =6.464$.}
\label{diagram}
\end{figure}

Fitting Eq.~\ref{eq:expfit} to all of the rod geometries tested allows us to map $\lambda(L_{*}, R)/d_\mathrm{L}$ as shown in Fig.~\ref{diagram}. Circular symbols denote regions of the $(L_*,R)$-space where trapping occurs with shading indicating the value of $\lambda/d_\mathrm{L}$, while triangles indicate regions without observable trapping, i.e., $N^\mathrm{trap}=0$ over the course of a simulation. The region of untrapped rods is bounded below by the spherical limit, i.e.\ $L_{*} = R^{-1}$ (black curve), on the left by the critical size ratio of rod diameter to bed particle diameter, $R = R_{t} = 6.464$ (vertical dashed line) and above by $L_{*}^{c}$ as determined by fits to Eq.~\ref{eq:criticalL} (red solid curve). Outside of this region, all particles eventually become immobilized. We note that this diagram corresponds to a disordered bed with a particular packing fraction of $\phi=0.60$. One would expect that increasing $\phi$ constricts the untrapped region, as denser packings reduce the effective pore size, trapping particles at lower $L_{*}$.

\subsubsection{Contacts and orientation of trapped rods}
\label{SecIIIA2}

To further elucidate the passing-trapping transition and the particle-level mechanisms underlying rod trapping, we quantify the ensemble averaged contact number of the trapped rods with the bed particles: $\langle Z \rangle = \sum_{i=1}^{N^{\mathrm{trap}}} Z_{i}/N^{\mathrm{trap}}$, where $Z_i$ is the number of contacts for rod $i$ and $N^\mathrm{trap}$ is the number of trapped rods. We find that each trapped rod has $Z=5$ so that $\langle Z \rangle = 5$, irrespective of $L_{*}$, $R$ and $n$, reflecting the minimum number of geometric constraints required to immobilize an elongated particle in the absence of friction. Furthermore and for the parameters investigated here, the trapped rod most frequently contacts four bed spheres and less frequently contacts three bed spheres.  This implies that at least one of the bed spheres is in double contact with the rod, a situation that is impossible for smooth rods where a bed sphere can at most have a single contact with the rod.  Consequently, we expect that smooth rods must be  longer than ``bumpy'' rods for trapping to occur.  

The trapped rod contact number of $\langle Z \rangle=5$ departs from the classical Maxwell criterion for isostaticity of a packing, which stipulates the number of contacts equals the number of degrees of freedom ($\langle Z \rangle = 6$ for the rod), because in the absence of friction a rod remains free to rotate about its long axis. Previous studies of isostatic beds of spherical particles indicate values for spheres of $\langle Z \rangle=6$ \cite{moukarzel1998isostatic, roux2000geometric, petit2025pressure}, although there are significant deviations from this value for nonspherical particles. For instance, the mean contact number $\langle Z \rangle$ for monodisperse rod packings is between about 5 and 10 for $A<40$ \cite{williams2003random, blouwolff2006coordination, wouterse2009contact}. In bidisperse rod packings, $\langle Z \rangle$ remains around 5 for $A<5$ \cite{marschall2018compression}. However, in rod-sphere mixtures, $\langle Z \rangle$ is as high as 10 at large rod concentrations \cite{anzivino2024shear}. Our result that isolated (intruder limit) rods are trapped below the expected isostatic limit suggests that their immobilization is dominated by the geometric entanglement and rotational constraints inherent to high-aspect-ratio bodies.

\begin{figure}[t]
\centering
\centering \includegraphics[scale=0.46]{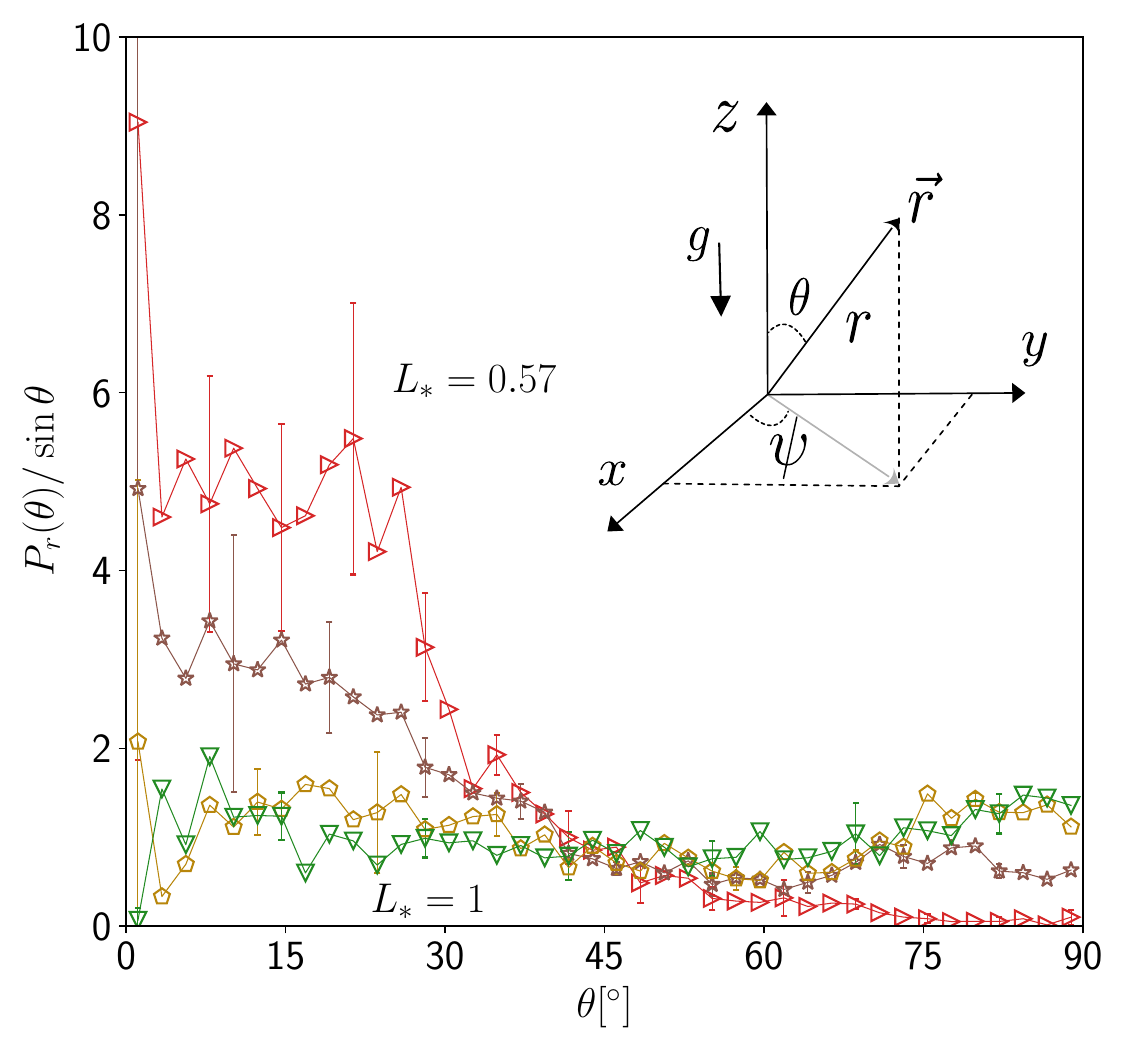}
\caption{Scaled polar density distributions of trapped rod orientation, $P_r(\theta)/\sin\theta$, vs polar angle, $\theta,$ at $R=7$ and different $L_{*} \in \{0.57, 0.71, 0.86, 1\}$. Data are averaged over five independent simulations with different initial bed sphere packings and rods positions. Error bars indicate the standard deviation and are only plotted for selected data points to improve readability.}
\label{trapped_orientation}
\end{figure}

Further insights into trapping can be gained by examining the orientation of trapped rods. Specifically, we calculate the polar, $ 0 \leq \theta \leq 180^{\circ}$, and azimuthal, $-180^{\circ} \leq \psi \leq 180^{\circ}$, angles using $\theta = \mathrm{arccos}\big(\Delta z / r \big)$ and $\psi = \mathrm{sgn}(\Delta y)\,\mathrm{arccos}\big(\Delta x / \sqrt{\Delta x^{2} + \Delta y^{2}} \big)$, where $r = \sqrt{\Delta x^{2} + \Delta y^{2} + \Delta z^{2}}$ is the distance between any two constituent spheres of the rod. The function $\mathrm{sgn}(\Delta y)$ provides the appropriate sign to ensure the correct quadrant. A sketch defining these angles along with the polar orientation distribution are shown in Fig.~\ref{trapped_orientation}. The rod azimuthal density distribution $P_r(\psi)$, is statistically uniform (not shown), confirming that trapped rods orient without a preferred horizontal direction. The rod polar density distribution defined by $P_r(\theta) = \int_{0}^{2\pi} P_{r}(\theta,\psi)\,\sin\theta d\theta d\psi$, is scaled by $\sin\theta$ to remove the geometric distortion inherent to spherical coordinates. This scaling ensures that each region of solid angle is equally weighted regardless of its location and actual area. Therefore, its normalization is $\int_{0}^{\pi} P_{r}(\theta)\,d\theta = 1$. The angle $\theta$ is restricted to $[0, 90^{\circ}]$ since the rod alignment at $0$ is equivalent to $180^{\circ}$. 

Figure~\ref{trapped_orientation} shows $P_r(\theta)/\sin(\theta)$ of trapped rods for different $L_{*}$ values at $R = 7$. For $L_{*} = 0.57$, rods are more frequently aligned vertically $(0 \leq \theta \leq 30^{\circ})$, while orientations with larger $\theta$ occur less frequently. As $L_{*}$ is increased, vertical alignment becomes less pronounced, while a slightly enhanced tendency for more horizontal orientations ($\theta > 60^{\circ})$ appears. 

\begin{figure}[t]
\centering
\centering \includegraphics[scale=0.46]{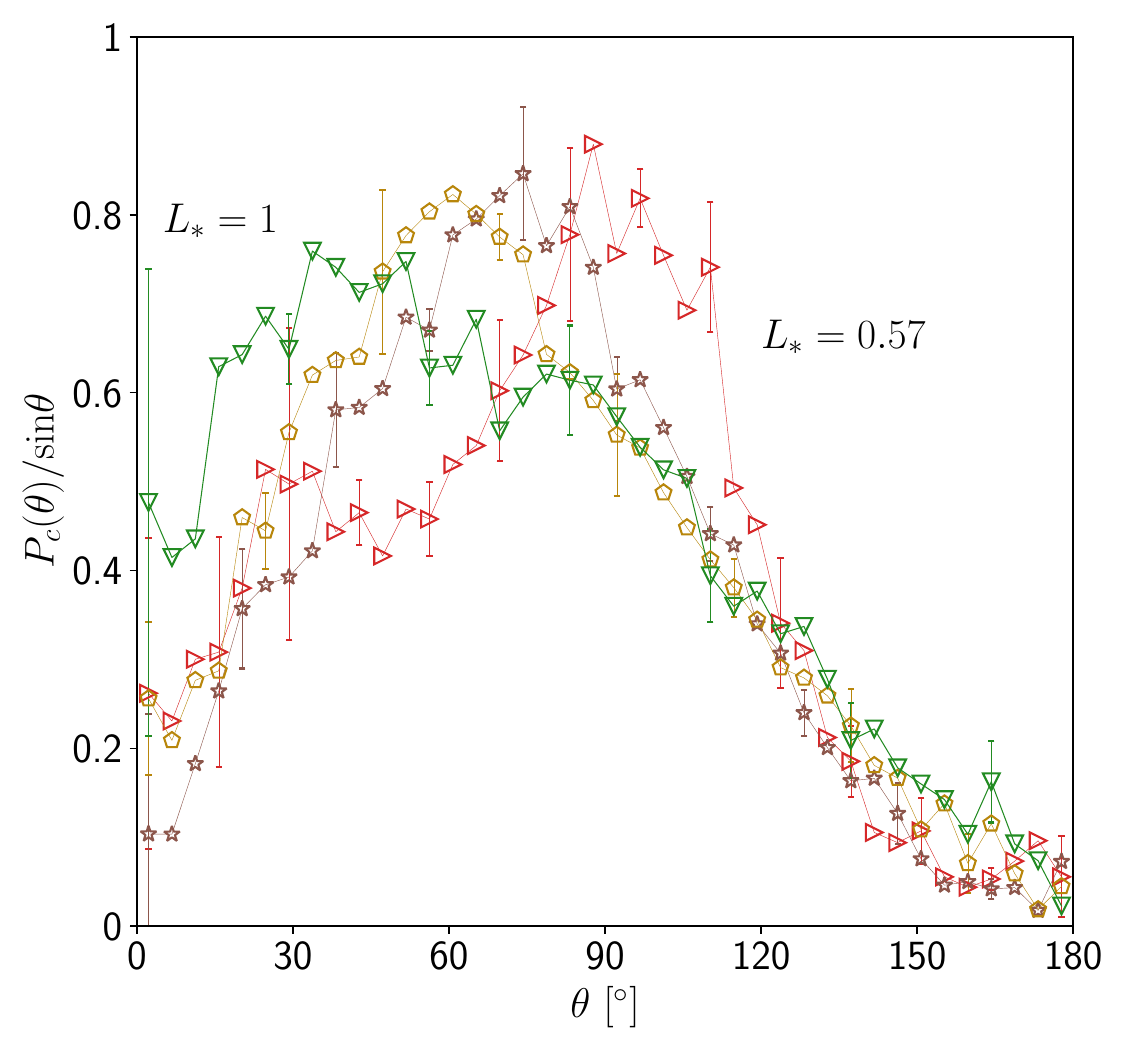}
\caption{Scaled polar density distributions of contact location, $P_c(\theta)/\sin\theta$, vs polar contact angle between trapped rods and bed spheres ($R = 7$) at different $L_{*} \in \{0.57, 0.71, 0.86, 1\}$. Data are averaged over five independent simulations with different initial bed spheres and rods positions. Error bars indicate the standard deviation and are only plotted for selected data points to improve readability.}
\label{trappedcontact}
\end{figure}

Whether a rod continues to percolate or becomes trapped also depends on the location of its contacts with the bed spheres, since these contacts dictate the mechanical constraints experienced by the rod that prevent it from percolating further. For this reason, we compute the angular distribution of contacts, $P_c(\theta)/\sin\theta$ with $\int_{0}^{\pi} P_{c}(\theta)\,d\theta = 1$, of trapped rods on the bed spheres in Fig.~\ref{trappedcontact} for the same values of $L_{*}$ shown in Fig.~\ref{trapped_orientation}. Here, $\theta$ represents the polar angle for the contact on the bed sphere with any constituent small spheres of the rod, measured with respect to the vertical axis. Figure~\ref{trappedcontact} displays a peak near $\theta \approx 90^{\circ}$ for $L_{*} = 0.57$, indicating that equatorial contacts are dominant for this rod length. The peak is shifted to lower angles as the rod length is increased. By considering both these contact distributions and the trapped-rod orientation data in Fig.~\ref{trapped_orientation}, a coherent picture of the trapping mechanisms emerges. Short rods $(L_{*} = 0.57)$ typically become arrested near the large sphere equator while remaining nearly vertically aligned. In contrast, the longest rods $(L_{*} = 1)$ are trapped with contacts located in a broader geometric zone between the pole and the equator and without any significantly preferred orientation.

\subsubsection{Stopping behavior}

\begin{figure}[t]
\centering \includegraphics[scale=0.58]{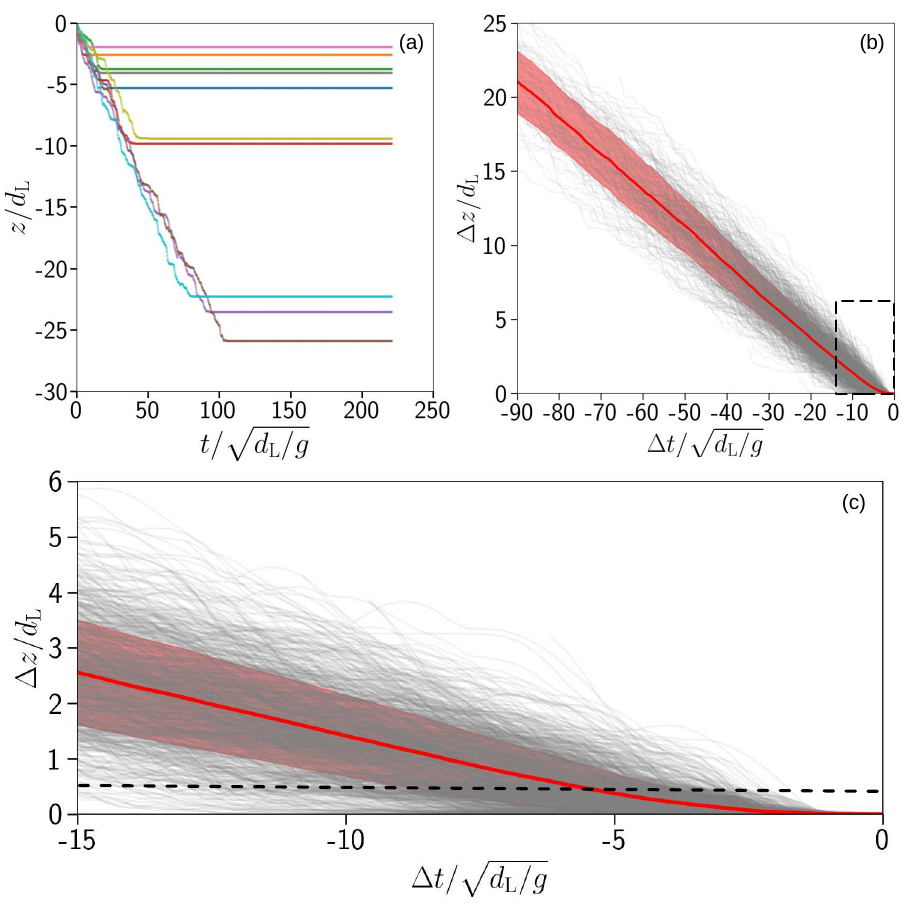}
\caption{(a) Scaled vertical $z$-positions of selected rod-like particles vs scaled time, $t/\sqrt{d_\mathrm{L}/g}$. (b) Average (red line) and individual (gray lines) scaled rod $z$-position relative to stopping depth, $\Delta z/d_\mathrm{L}$ vs scaled time to stopping time, $\Delta t/\sqrt{d_\mathrm{L}/g}$ for $R = 7$ and $L_{*} = 0.71$. Red shaded area indicates $1\sigma$ and characterizes the dispersion of the trajectories. Dashed square box indicates region displayed in (c). (c) Average trajectory shows deceleration of rods, indicating a stopping distance of $s_{d} \approx 0.5 d_\mathrm{L}$ (dashed horizontal line).} 
\label{TrapRegime}
\end{figure}

As rod-like particles fall through the interstitial spaces of a disordered bed of large spheres, they either pass through or become stuck within the pore structure depending on the values of $L_{*}$ and $R$ as discussed above. Here we consider the transition in kinematics of rods as they become trapped, i.e.\ as their motion is arrested. First consider rod trajectories for $R = 7$ and $L_{*} = 0.71$, a case where the rod eventually becomes trapped. Figure~\ref{TrapRegime}(a) shows an initial linear decrease in position of the midpoint of the rod with time for rods with several different initial conditions, indicating an average velocity that is nearly the same for all initial conditions shown. The eventual development of plateaus for all trajectories indicates the immobilization of the rods at their final trapping depth. In all cases, a pronounced deceleration occurs immediately prior to trapping. 

To characterize the spatial extent of this rapid deceleration, we compute an ensemble-averaged trajectory over different rod initial conditions, providing a statistically representative profile of the rod position as it approach immobilization, similar to the approach previously used for percolation of spherical particles~\cite{gao2023percolation}. The trajectories are aligned to a common time and $z$-position at which the particle becomes trapped (as defined in Sec.~\ref{SecIIIA1} with the exception that the mean velocity over an averaging interval of $1.2 \sqrt{d_{\mathrm{L}}/g}$ is used in place of the instantaneous velocity) and indicated by where the trajectory abruptly flattens in Fig.~\ref{TrapRegime}(a). The resulting times and positions are used to align the trajectories of all rods so that the shifted time and location of immobilization correspond to $(\Delta t, \Delta z)=(0,0)$ for every rod. 

The average trajectory along with all of the individual rod trajectories are displayed in Fig.~\ref{TrapRegime}(b, c) for the same $R=7$ and $L_{*}=0.71$ presented in Fig.~\ref{TrapRegime}(a). On average, rods display a linear displacement with time far above the stopping point, corresponding to a constant percolation velocity with relatively small variability (shaded area). As the rods approach the stopping point, they decelerate and come to rest at $(\Delta t, \Delta z) = (0, 0)$. The stopping distance $s_{d}$, defined as the average $z$-position at which rods begin to decelerate based on the average trajectory deviating from being linear, is $s_{d} \approx 0.5 d_{\mathrm{L}}$ [dashed line in Fig.~\ref{TrapRegime}(c)]. This value is similar to the stopping distance observed for spheres with $R = 5$ \cite{gao2023percolation}. In both cases, percolating spheres and rods slow down and stop within a distance less than the diameter of a bed sphere. Consistent trends are observed across the other $R$ and $L_{*}$ values studied. However, identifying the precise stopping point for large $R$ and $L_{*}$ cases is challenging as the exact moment of arrest is obscured by the increased occurrence of slow dynamics prior to trapping. 

\subsection{Untrapped rods}
\label{SecIIIB}
In this section, we study the behavior of rods \emph{while they are percolating} regardless of whether they eventually become trapped or not. In particular, we characterize the mean percolation velocity, rod orientation during percolation, and collision location on the bed spheres as functions of the parameters defining the rod geometry. 

\subsubsection{Percolation velocity}

\begin{figure}[t]
\centering \includegraphics[scale=0.48]{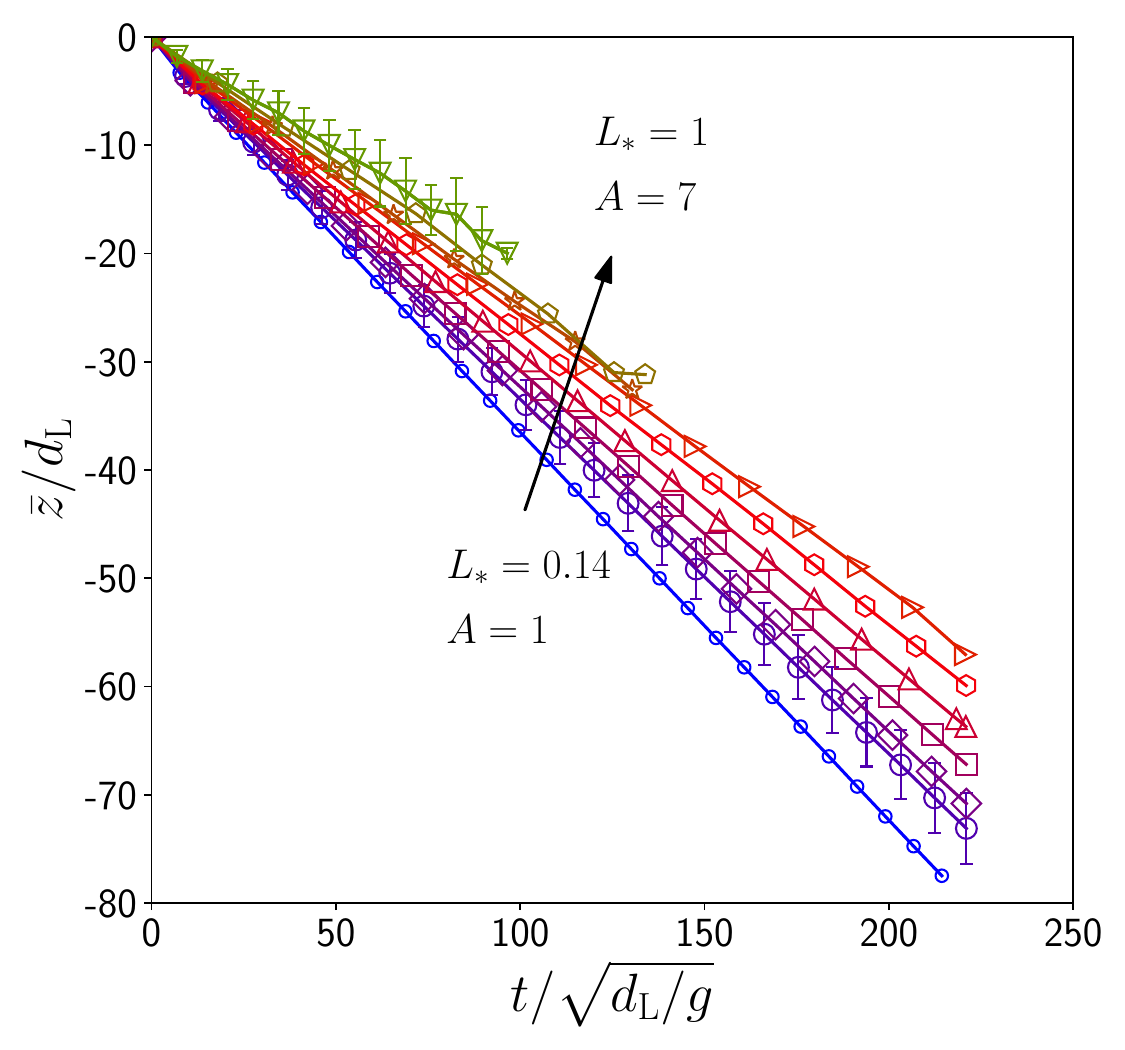}
\caption{Average scaled rod displacement, $\bar{z}/d_{\mathrm{L}}$, vs scaled time, $t/\sqrt{d_{\mathrm{L}}/g}$, for size ratio $R=7$ at different scaled lengths $L_{*} \in \{0.14, 0.21, 0.29, 0.36, 0.43, 0.50, 0.57, 0.71, 0.86, 1\}$ and associated aspect ratios $A=l/d_\mathrm{S}=L_*R$. Note that $L_{*}=0.14$ corresponds to spherical particles ($A=1$) since $L_*=A/R=1/R$. Slopes of lines correspond to the dimensionless percolation velocity, $\tilde{v}_p$. Error bars indicate the standard deviation and are plotted for selected rod geometries to improve readability.}
\label{fig2}
\end{figure}

The average vertical percolation velocity, $v_p$, is determined from the slope of the average vertical displacement, $\bar{z}$, over an ensemble of untrapped rods after they enter the bed as a function of time. The vertical position of each rod is tracked to the end of the simulation or until the time where the rod is trapped. Figure~\ref{fig2} shows the scaled average vertical displacement, $\bar{z}/d_\mathrm{L},$ as a function of the scaled time, $t\sqrt{g/d_{\mathrm{L}}},$ for $R = 7$, which is above the critical trapping size ratio $R_{t} = 6.464$, and for various lengths $L_{*}$ and corresponding aspect ratios $A$. The slope is the dimensionless percolation velocity, $\tilde{v}_{p} = v_{p}/\sqrt{gd_{\mathrm{L}}},$ which decreases as $A$ and $L_{*}$ increase. The scaling factor $\sqrt{gd_{\mathrm{L}}}$ reflects the characteristic free fall velocity of a particle over a distance $d_{\mathrm{L}}$. The percolation velocity of short rods is only slightly slower than that of spheres ($L^*=0.14$), and the percolation velocity decreases as the length of the rod increases. Note that for $L_{*} \geq 0.71$, the percolation of the rods is limited to shorter times due to their more rapid trapping. 

\begin{figure}[t]
\centering \includegraphics[width=0.9\columnwidth]{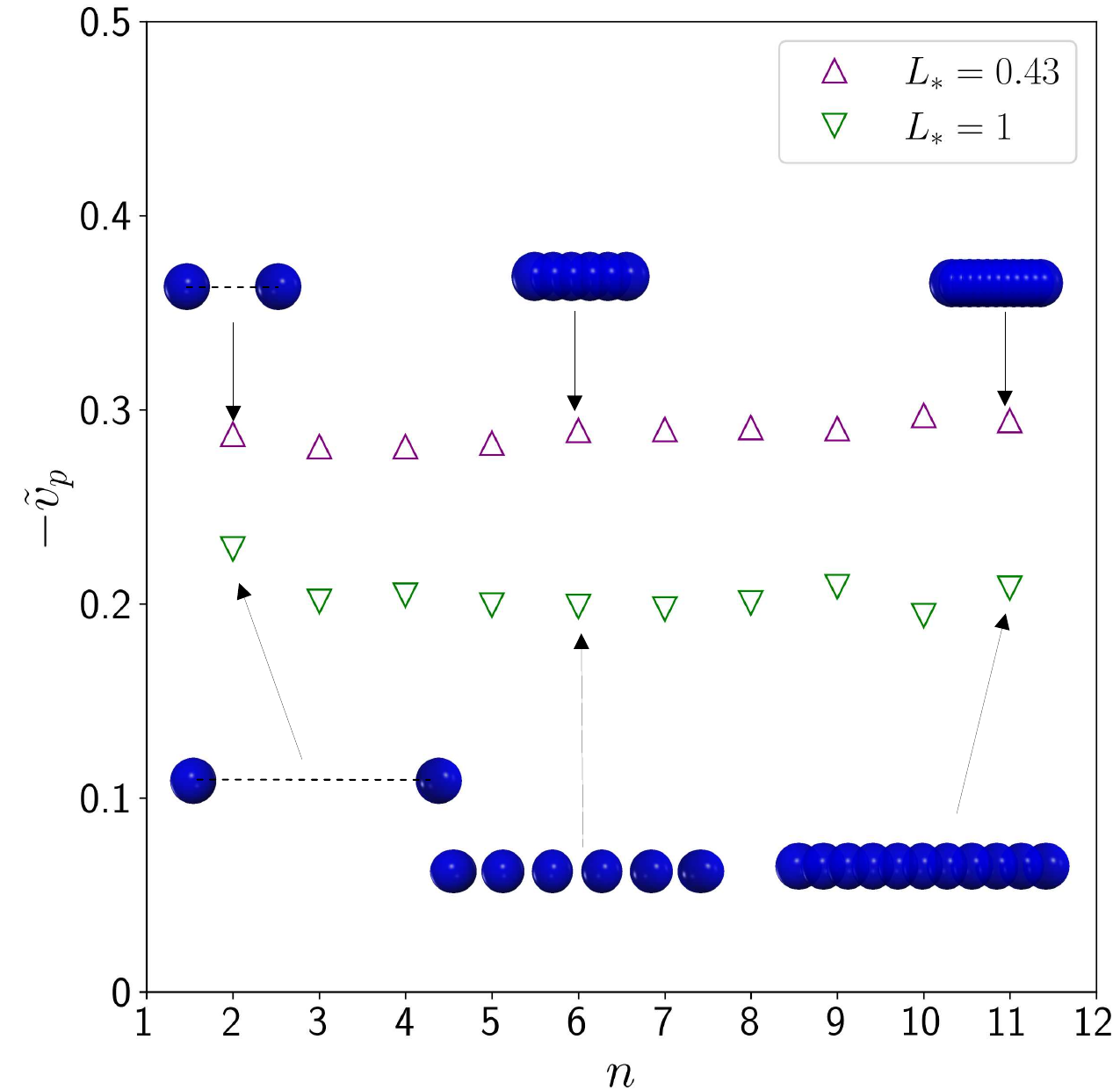}
\caption{Dimensionless percolation velocity, $\tilde{v}_{p}$, vs the number of constituent spheres, $n$, for two fixed values of scaled rod length $L_{*}$ at $R = 7$. }
\label{vp_constituents}
\end{figure}

At this point it is essential to demonstrate that the choice of rod-discretization, specifically the number of constituent spheres, $n$, does not inadvertently influence the results. As illustrated in Fig.~\ref{vp_constituents} for two fixed rod lengths at $R=7$, the dimensionless percolation velocity, remains effectively constant as $n$ is varied. In these cases, the dimensionless rod length $L_{*}$ is held constant while the spacing and number of constituent spheres are varied. For both $L_{*} = 0.43$ and $L_{*} = 1$, $\tilde{v}_{p}$ remains approximately constant at $0.28$ and $0.2$, respectively. The result that $\tilde{v}_{p}$ is independent of $n$ confirms that the percolation dynamics for these frictionless rods is governed by the rod geometry (i.e., length and diameter) and is independent of the resolution of the multi-sphere model.

\begin{figure}[t]
\centering \includegraphics[scale=0.47]{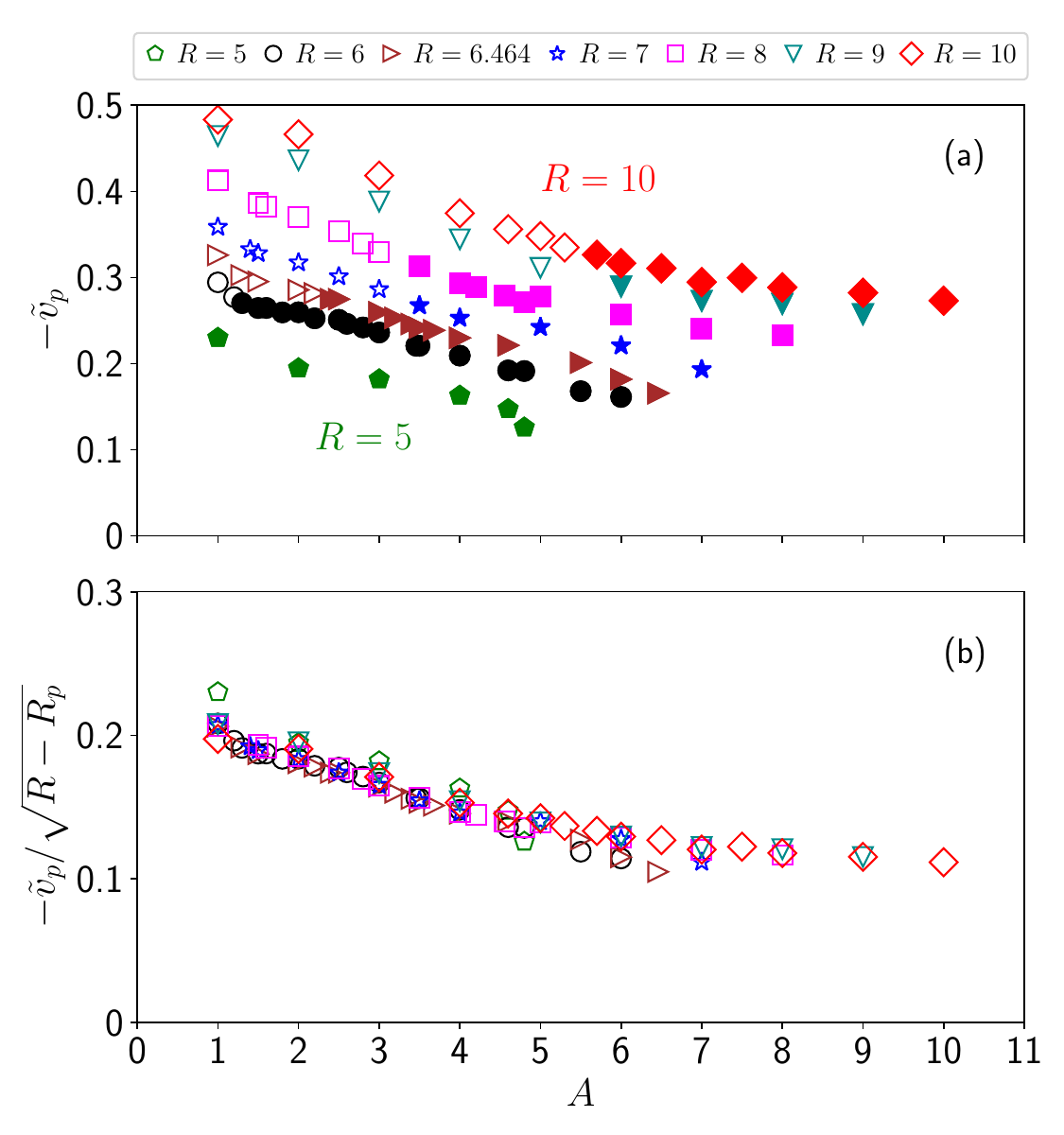}
\caption{ (a) Dimensionless rod percolation velocity, $\tilde{v}_{p} = v_{p}/\sqrt{gd_{\mathrm{L}}}$, vs length to diameter aspect ratio, $A$, at different size ratios, $R$. Open symbols correspond to parameters where rods are never trapped; filled symbols indicate $\tilde{v}_p$ measured up to the point where rods are trapped. (b) $\tilde{v}_{p}/\sqrt{R-R_{p}}$ vs $A$ where $R_p=4.5$ is the ratio between the average pore diameter and the rod diameter (see text), with open symbols for both trapped and untrapped cases.}
\label{vel_w}
\end{figure}

The dimensionless percolation velocity obtained by fitting a line to each data series over all simulation conditions is shown as a function of $A$ at different $R$ in Fig.~\ref{vel_w}(a). The percolation velocity decreases with increasing $A$ for fixed $R$. The lower velocity of longer rods arises from the need to satisfy a smaller range of acceptable orientation angles before they can pass through the pores formed by the large bed particles. Figure~\ref{vel_w}(a) includes data for $R < R_{t}$. As $R$ is decreased, it becomes increasingly unlikely for a rod to encounter pore throats significantly larger than the rod diameter, thereby slowing percolation. At larger $R$, the rod diameter is small enough for the rod to easily pass through the pore throats, resulting in higher velocities. The values for the percolation velocity corresponding to spheres ($A=1$) of $0.25 \leq v_p \leq 0.36$ for $5 \leq R \leq 7$ in Fig.~\ref{vel_w}(a) are about $50\%$ higher than those found previously for spherical particles
\cite{gao2023percolation} with a higher restitution coefficient of 0.8. This is expected as higher restitution coefficients can significantly reduce $v_p$ due to increased velocity fluctuations which frustrate percolation~\cite{vyas2026fine}.

In Fig.~\ref{vel_w}(a) we distinguish between geometries for which rods eventually become trapped (filled symbols) from those where the rods never become trapped (open symbols). There is no sharp change in the percolation velocity between rods that eventually become trapped and rods that never become trapped. Hence, the propensity for trapping does not significantly alter the dynamics of the rods while they are percolating, consistent with results for spherical particles \cite{gao2023percolation}.

The percolation velocities as a function of $A$ for rods with different $R$ collapse when scaled by $\sqrt{R - R_{p}}$, as shown in Fig.~\ref{vel_w}(b), where $R_{p}$ is indicative of the mean pore diameter of the large particle bed $(\phi=0.6)$. The slope of the scaled percolation velocity vs $A$ is significantly steeper for $A<5$ and tends to flatten for larger $A.$  Although the $\sqrt{R-R_p}$ scaling is based on a heuristic argument related to the clearance between a fine spherical particle and the mean pore diameter (related to the mean free path)~\cite{vyas2026fine}, it collapses the rod data surprisingly well, most likely because it reflects the clearance available for a rod to pass lengthwise through a pore throat.  The $R=5$ case deviates slightly from this scaling, which is likely attributable to the proximity of this size ratio to the parameter $R_{p}=4.5$ utilized in the scaling model (i.e., the collapse of the data at small $R$ is more sensitive to the exact value of $R_p$).  We note that the value of $R_p=4.5$ used here is slightly larger than the value of $R_p=4$ used in \cite{vyas2026fine}.

\begin{figure}[t]
\centering \includegraphics[scale=0.47]{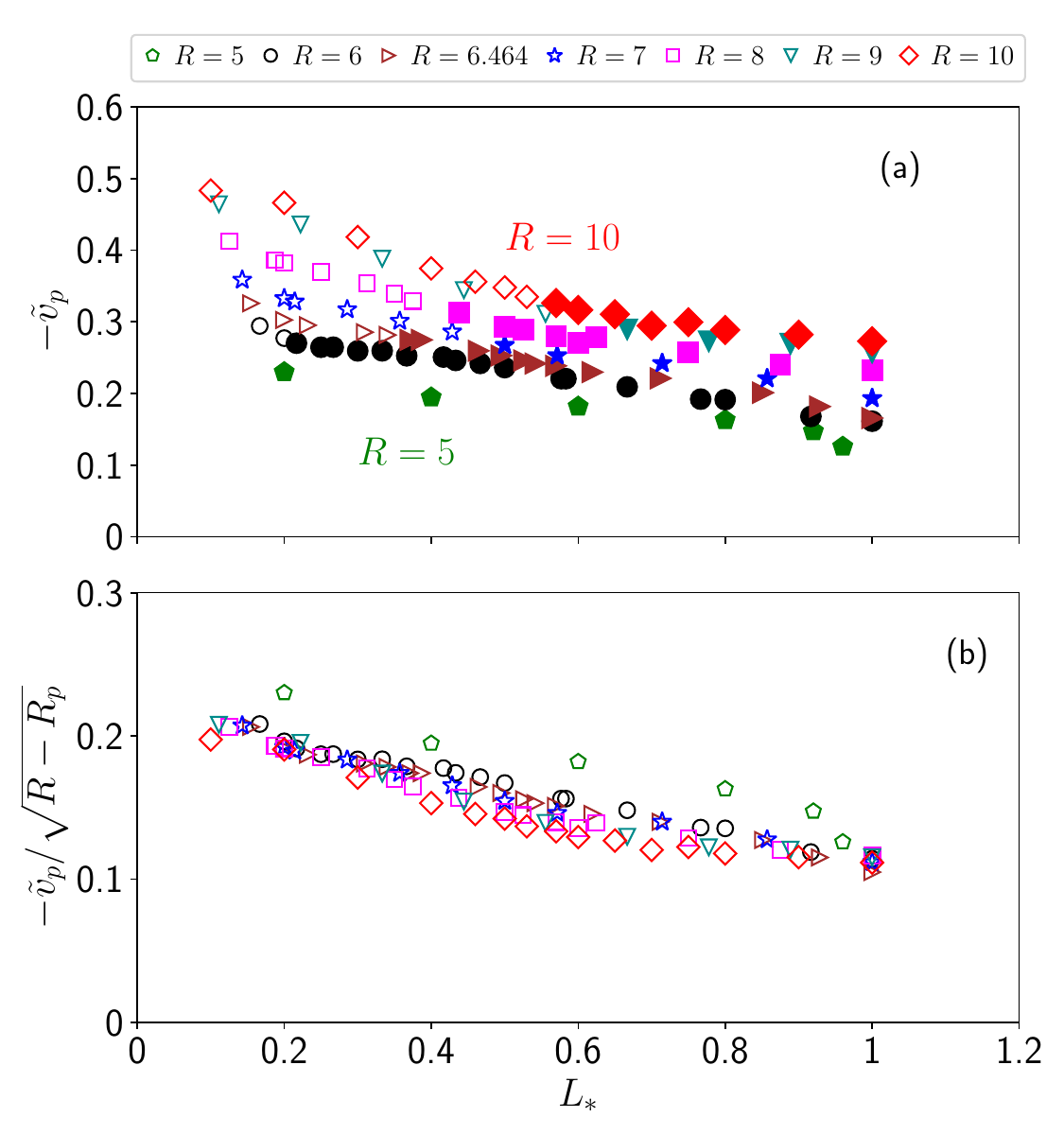}
\caption{ (a) Dimensionless rod percolation velocity, $\tilde{v}_{p} = v_{p}/\sqrt{gd_{\mathrm{L}}}$, vs scaled rod length, $L_{*}$, at different size ratios $R$. Open symbols correspond to parameters where rods are never trapped; filled symbols indicate $\tilde{v}_p$ measured up to the point where rods are trapped. (b) $\tilde{v}_{p}/\sqrt{R-R_{p}}$ vs $L_{*}$ with $R_p=4.5$ and open symbols for both trapped and untrapped cases.}
\label{vel_lstart}
\end{figure}

Since $R$, $A$ and $L_{*}$ are not independent, i.e., $L_{*}=A/R$, we additionally plot $\tilde{v}_p$ as a function of $L_{*}$ for various $R$ in Fig.~\ref{vel_lstart}(a). As would be expected, $\tilde{v}_p$ decreases with increasing $L_{*}$, since rods with lengths approaching the size of the bed particles are more constrained in passing through the interstices in the packed bed. Again, scaling $\tilde{v}_p$ by $\sqrt{R-R_p}$ with $R_p=4.5$ collapses the data well, as shown in Fig.~\ref{vel_lstart}(b), except for the $R=5$ data, which is again a noticeable outlier because $R=5$ is much smaller $R_t$. The decrease in the scaled percolation velocity is approximately linear with $L_*$. 

\begin{figure}[t]
\centering \includegraphics[scale=0.54]{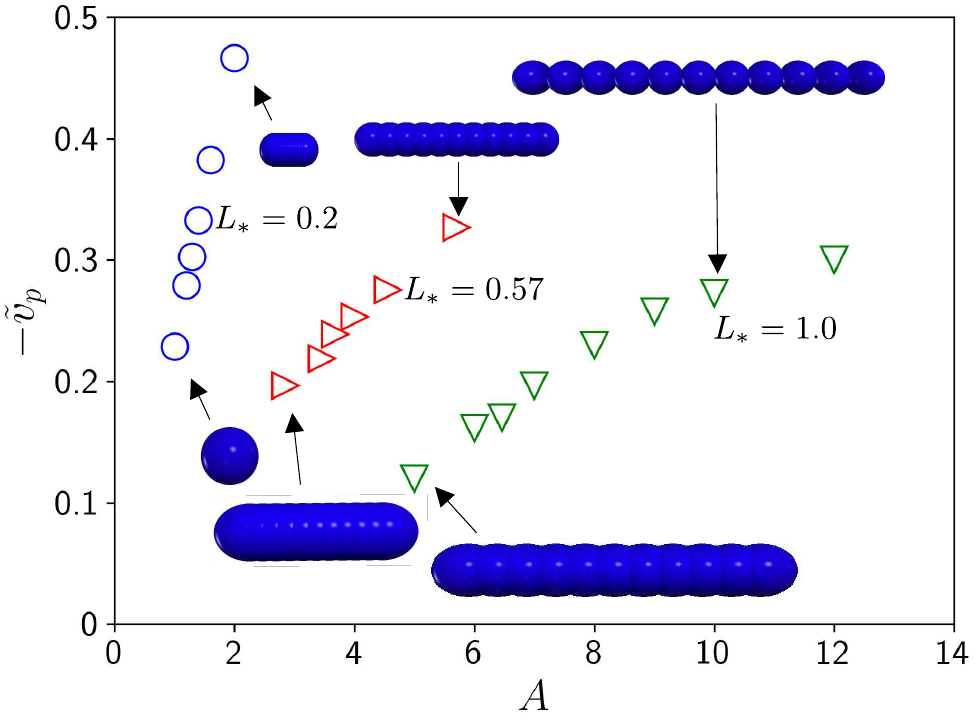}
\caption{Scaled percolation velocity, $\tilde{v}_p$, vs aspect ratio, $A$, for three rod lengths, $L_{*}$. Note that $R$ at constant $L_*$ varies with $A$ as $R = L_{*}/A$.}
\label{l*constant}
\end{figure}

Figure~\ref{l*constant} shows $\tilde{v}_p$ as a function of $A$ for three different values for $L_{*}$. For the longest rods considered, $L_{*} = 1,$ having a length corresponding to the diameter of a bed sphere $d_{\mathrm{L}}$, $\tilde{v}_p$ increases monotonically with $A$, reflecting the dependence on the size ratio $R = A/L_{*}$. Thus for rods of fixed length, thinner rods percolate faster than thicker ones. As the rod length is decreased, shorter rods consistently exhibit higher $\tilde{v}_p$ than longer rods while preserving the same qualitative dependence on $A$. This behavior is expected since shorter rods experience fewer and less persistent contacts with the surrounding bed spheres during rotational motion as they tumble between the bed particles, reducing geometric restrictions and allowing for more efficient downward transport through the bed.

\begin{figure}[t]
\centering \includegraphics[scale=0.42]{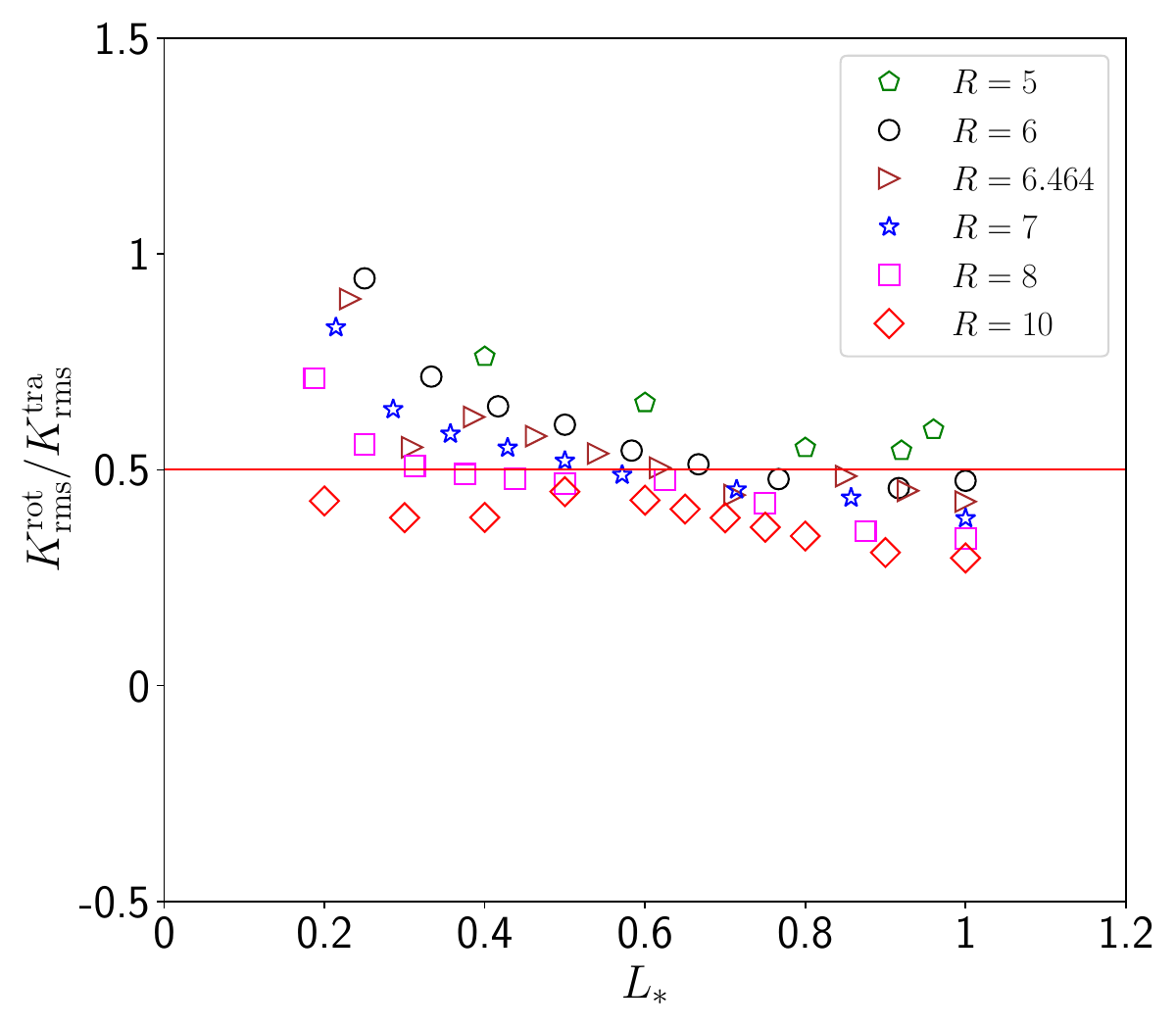}
\caption{Ratio of the root mean square (RMS) rotational kinetic energy to translational kinetic energy, $K_{\mathrm{rms}}^{\mathrm{rot}}/K_{\mathrm{rms}}^{\mathrm{tra}}$, vs dimensionless rod length, $L_{*}$, at different $R$ values. The red horizontal line corresponds to $K_{\mathrm{rms}}^{\mathrm{rot}}/K_{\mathrm{rms}}^{\mathrm{tra}}= 0.5$.} 
\label{vp_omega_rms}
\end{figure}

To gain further insight into the particle dynamics during percolation, we measure the ratio of the RMS rotational and translational kinetic energies with respect to the center of mass of the rod: $$K_{\mathrm{rms}}^{\mathrm{rot}}/K_{\mathrm{rms}}^{\mathrm{tra}} = (L_{*}^{2}d_{\mathrm{L}}^{2}\omega^{2}_{\mathrm{rms}}/6v^{2}_{\mathrm{rms}})^{2} [1 + 3/(4R^{2}L_{*}^{2})],$$ where $\omega_{\mathrm{rms}}$ and $v_{\mathrm{rms}}$ are the root mean square rotational and translational velocities, respectively. Figure \ref{vp_omega_rms} plots this ratio as a function of $L_{*}$ for various $R$. The translational excitation remains consistently higher than its rotational counterpart, with the ratio $K_{\mathrm{rms}}^{\mathrm{rot}}/K_{\mathrm{rms}}^{\mathrm{tra}} < 1$. For short rods ($0.15 \leq L_{*} < 0.3$), rotational excitation is nearly comparable to translational excitation, suggesting an increase of tumbling behavior and a near equipartition of energy. However, as rod length increases, this ratio decreases significantly as the extended geometry restricts rotational freedom within the pore network. Similarly, $K_{\mathrm{rms}}^{\mathrm{rot}}/K_{\mathrm{rms}}^{\mathrm{tra}}$ diminishes as the size ratio $R$ increases, indicating that translational fluctuations becomes increasingly dominant for high-aspect-ratio rods.

\subsubsection{Orientation and collision location of untrapped rods}
\label{SecOrient}

Unlike spheres, rods have distinct sets of symmetry axes that continuously change their orientation as the rod percolates through the bed. To quantify the rod orientation during percolation, we calculate the polar angle of the rod's long axis, $0 \leq \theta \leq 90^\circ$, as defined in Sec.~\ref{SecIIIA2}, from the time that the rod's center of mass is $d_\mathrm{L}$ below the top of the bed until the end of the simulation or until the rod becomes trapped. Untrapped rods are identified by the condition $\left|v_{z}\right| > 2.2 \times 10^{-2} \sqrt{g d_{\mathrm{L}}} \approx 0.3$\,cm/s.

\begin{figure}[t]
\centering \includegraphics[scale=0.57]{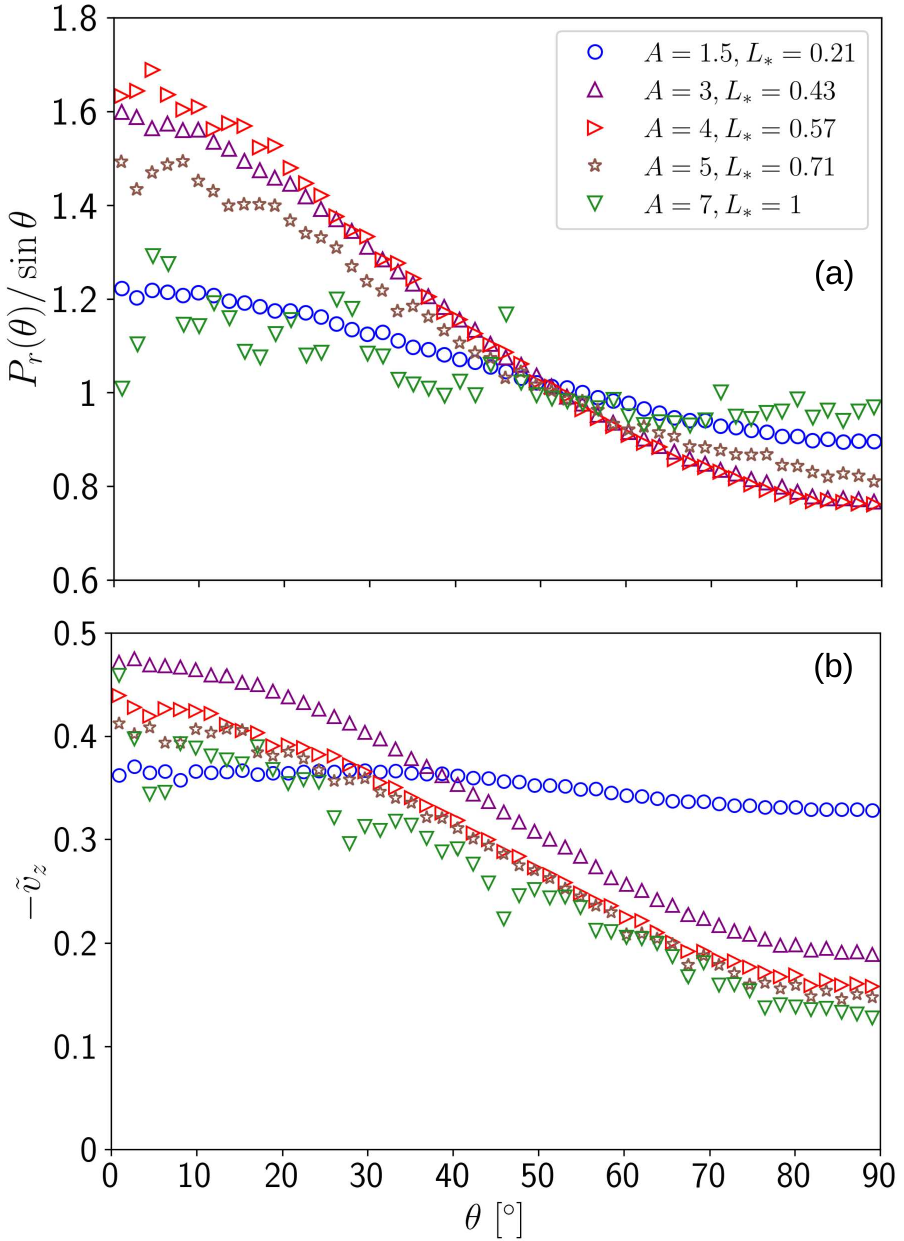}
\caption{(a) Orientation distribution, $P_r(\theta)/\sin\theta$, and (b) average vertical velocity $\tilde{v}_z$ vs $\theta$ for percolating rods with $R = 7$ at different $A$ and $L_{*}$. $N = 5000$ rods are used for increased statistical accuracy.}
\label{orien_dist}
\end{figure}

Figure~\ref{orien_dist}(a) shows the scaled orientation distribution, $P_r(\theta)/\sin\theta$, for percolating rods for several values of $A$ and $L_{*}$ with $R = 7$. Percolating rods have preferred orientations that depend on $L_{*}$. For $L_{*} = 0.21$, the long axis of short rods has a slightly greater tendency to align vertically $(\theta = 0)$ than horizontally $(\theta = 90^{\circ})$ (a tendency that vanishes in the spherical-rod limit $L_* \rightarrow 1/R$). Vertical alignment is enhanced as $L_*$ is increased with maximal vertical bias occurring in the vicinity of the critical length, $L_*\approx L_*^c \approx 0.52,$ separating the passing and trapping regimes, see Fig.~\ref{diagram}. As $L_*$ is increased beyond $L_*^c$, vertical bias diminishes and is nearly absent at $L_*=1.$ This behavior is consistent with the distribution of trapping orientations discussed in the context of Figs.~\ref{trapped_orientation} and \ref{trappedcontact}, indicating that trapping occurs without significant reorientation. Shorter rods tend to align vertically when passing through a pore and tend to be trapped in more vertical orientations. Longer rods display less tendency to orient in any particular polar orientation when passing through the bed and consequently have a lower tendency to trap in any particular orientation. The azimuthal distribution for $P_{r}(\psi)$ (not shown) is uniform and independent of $L_{*}$, indicating that the rods have no preferred orientation in the horizontal ($x$-$y$) plane as expected for a randomly packed bed.

To further examine rod percolation through the packed bed, we plot the average vertical velocity, $\tilde{v}_z$, as a function of $\theta$ in Fig.~\ref{orien_dist}(b). In this manner, we can determine the angular dependence of the rod velocity, revealing how orientation influences transport speed through the interstitial spaces. Note that $\tilde{v}_z$ can be integrated over $P_{r}(\theta)$ to obtain the percolation velocity $\tilde{v}_p$. For $L_{*} = 0.21$, the vertical velocity depends only weakly on orientation, indicating that short rods descend through the bed without a preferred orientation and maintain an almost uniform vertical velocity. When the rod length is increased to $L_{*} = 0.43$, orientation effects become pronounced: rods aligned closer to the vertical direction exhibit a substantial increase in $\tilde{v}_z$, whereas horizontally aligned rods descend more slowly. The effect is similar for longer rods, but the greater rod length slightly reduces $\tilde{v}_z$ across all $\theta$, as it hinders passage through the pore throats formed by the bed spheres. 

\begin{figure}[t]
\centering \includegraphics[scale=0.45]{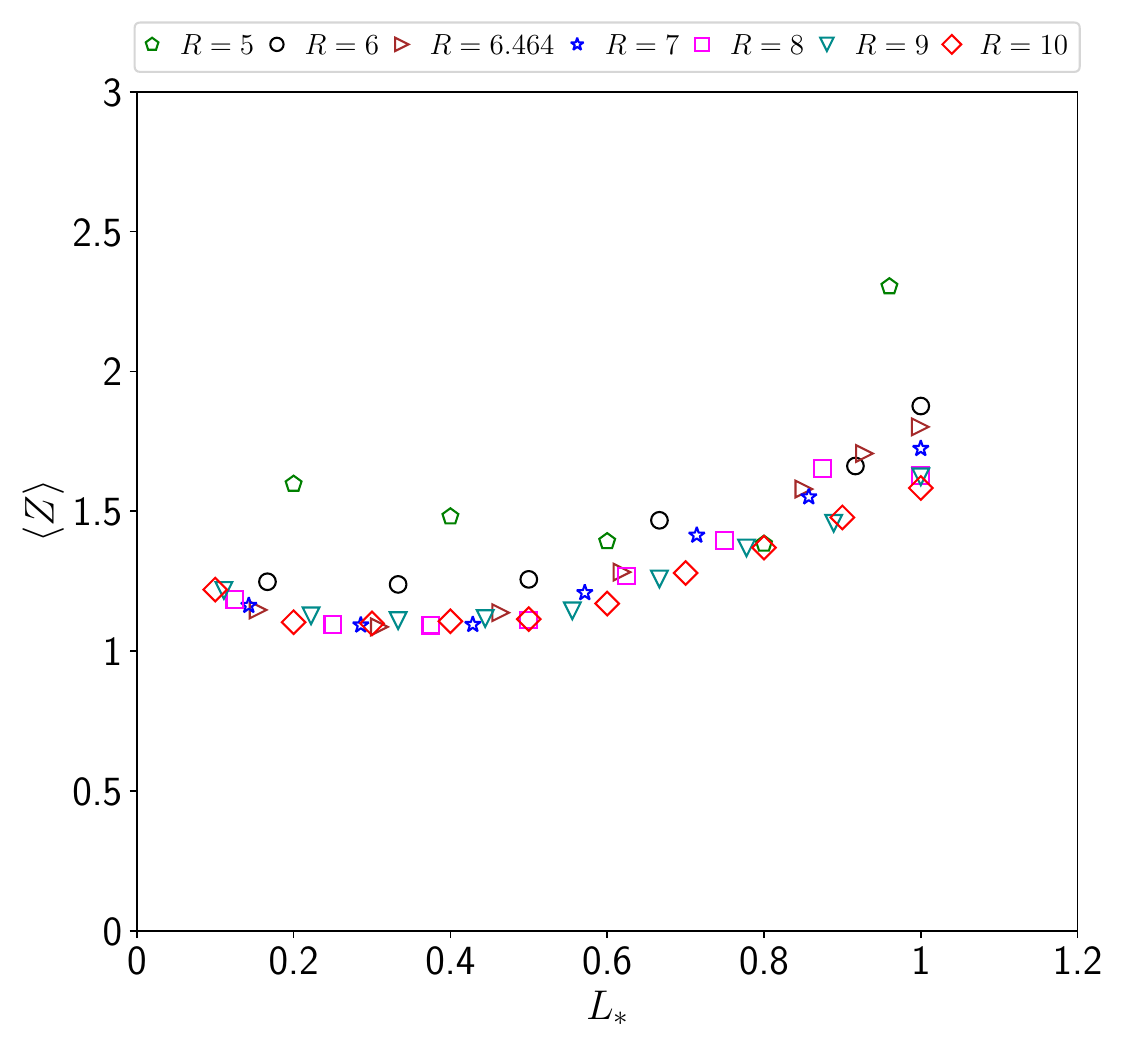}
\caption{Mean contact number, $\langle Z \rangle$, vs $L_{*}$ for percolating rods at different $R$.}
\label{menaz_untrapping}
\end{figure}

The differences in percolation velocity and orientation as rod length increases suggest that rods interact with the spheres in the packed bed in different ways during percolation. To elucidate this behavior, we analyze the contacts formed by percolating rods after they enter the sphere bed, considering only untrapped rods. All rod-sphere contacts are tracked over the entire duration of the simulation. Here, rod contacts are defined as the contacts made by one or more of the constituent small spheres composing the rod with any bed spheres. As shown in Fig.~\ref{menaz_untrapping}, the average instantaneous mean contact number $\langle Z \rangle$ is between 1 and 2 for percolating rods. It has a broad minimum near 1 for $0.2 \leq L_{*} \leq 0.4$. It increases slightly for smaller $L_{*}$ and nearly doubles for the longest $L_{*}$ compared to the shortest $L_*$. Except for the $R=5$ and $6$ cases, which are expected to have larger values for $\langle Z \rangle$ simply because the rod diameter exceeds the minimum pore throat diameter, the data for all $R$ have similar numbers of contacts as a function of $L_{*}$. However, at $L_{*}=1$ a distinct ordering of $\langle Z \rangle$ with $R$ is evident where $\langle Z \rangle$ decreases monotonically with increase $R$. Hence, for cases where $R > R_t$, rod length drives the number of contacts with little influence of rod diameter, while both rod length and diameter play roles as the length of the rod approaches the size of the bed spheres.

\begin{figure*}[t]
\centering
\centering \includegraphics[width=0.9\textwidth]{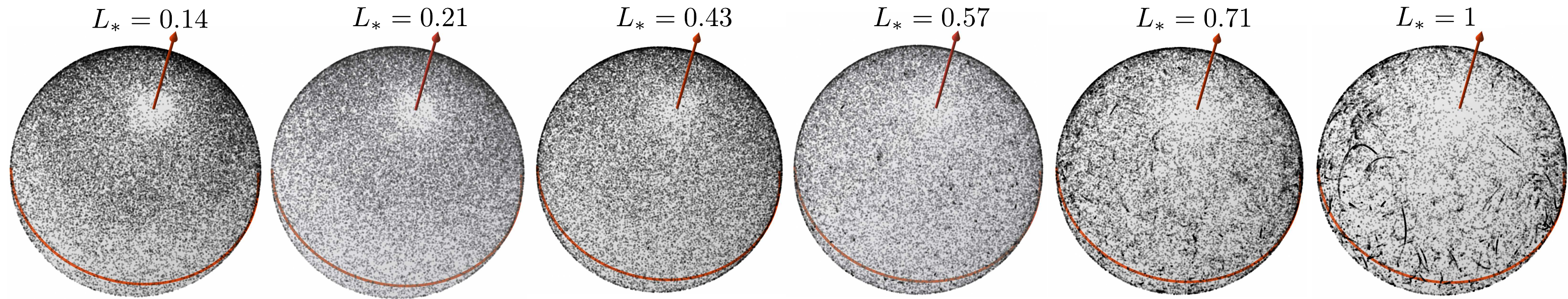}
\caption{Contact locations of percolating rods overlaid on a single bed sphere for different rod lengths $L_{*}$ and $R= 7$. Symbols are semi-transparent making regions with overlapping symbols darker. To enhance interpretability, we show only a representative subset of contact points, maintaining a consistent count across all cases. Red curves represent the equator of the sphere (horizontal plane), while red arrows indicate the positive $z$-direction (vertical). $N = 5000$ rods are used to improve statistical accuracy.}
\label{CPoints}
\end{figure*}

\begin{figure}[t]
\centering \includegraphics[scale=0.4]{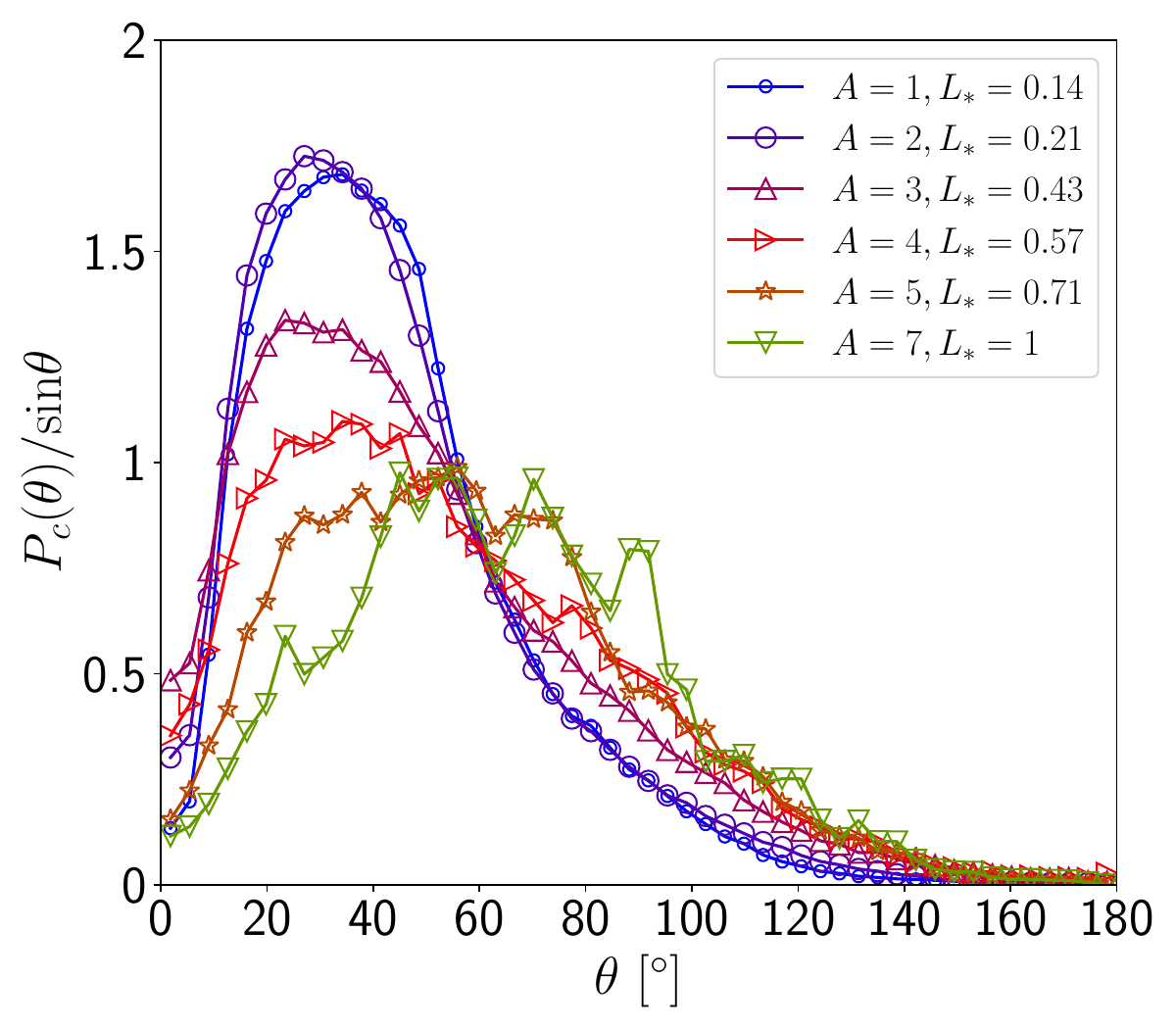}
\caption{Scaled polar contact angle distributions of percolating rods, $P_c(\theta)/\sin\theta$, for different rod lengths $L_{*}$ and aspect ratios $A$ with $R=7$. $N = 5000$ rods are used to improve statistical accuracy.} 
\label{CDist}
\end{figure}

Figure~\ref{CPoints} displays contact points of the percolating rods with bed spheres overlaid onto a single bed sphere at different $L_{*}$. The spheres are tilted slightly (noting the equator and $z$-axis in red) to more clearly display the contact points. A quantitative representation of the contact locations in Fig.~\ref{CPoints} is presented in Fig.~\ref{CDist} for different rod lengths using the scaled contact polar distribution, $P_c(\theta)/\sin\theta$, noting that $P_c(\psi)$ is uniform (not shown). What is particularly striking in Fig.~\ref{CPoints} is that there are few contacts near the poles of the bed spheres, regardless of $L_{*}$. This is a consequence of bed spheres being shielded from contacts with rods by other bed spheres just above them. This shielding effect is evident in the low values of $P_c(\theta)/\sin(\theta)$ for small $\theta$ in Fig.~\ref{CDist}. This effect is strongest for spherical fine particles $(L_{*}=0.14)$ where most fine spherical particles contact the bed spheres in the upper hemisphere, mostly in the range $20^{\circ} \leq \theta \leq 60^{\circ}$. Contacts in the lower hemisphere ($\theta > 90^{\circ}$) are primarily the result of scattering events as the fine particles ricochet off nearby large spheres. With increasing rod length, the contacts distribute more broadly toward the equatorial region of the upper hemisphere and into the lower hemisphere. This dispersion reflects the stronger orientational constraints experienced by longer rods as they approach and interact with the sphere bed during their downward motion. In particular, for $L_{*} = 1$, where the rod length equals the diameter of the large spheres, most contacts occur in the equatorial region. Moreover, as is evident in Fig.~\ref{CPoints}, long rods (here $L_* \in \{0.71, 1\})$ can exhibit continuous contact lines on the bed spheres, indicative of persistent contacts as a long rod slides down the surface of a bed sphere. This behavior further demonstrates that longer rods experience increased geometric frustration traversing the packed bed with low velocity [(see Fig.~\ref{vp_omega_rms} and Fig.~\ref{orien_dist}(b)], leading to longer contact durations and interactions over a broader range of surface locations.

\section{Conclusion}

We investigate percolation of single rod-like particles through a static disordered bed of large spheres, demonstrating how particle anisotropy alters the percolation behavior relative to the more studied case of spherical particles. By systematically varying the rod aspect ratio $A$, relative length $L_{*}$, and size ratio $R$, we quantify and examine two distinct regimes—passing and trapping—governed by the interplay between pore geometry, rod orientation, and contact mechanics.

Untrapped rods percolate with a constant terminal velocity that depends on $L_{*}$ and $R$. Remarkably, the data collapse when normalized by the velocity scale $\sqrt{g d_{\mathrm{L}}(R - R_{p})}$. This scaling unifies sphere and rod dynamics, thereby extending percolation models for spheres to rod-like particles. Furthermore, the decrease in velocity with increasing $L_{*}$ is approximately linear, reflecting the progressively tighter geometric constraints imposed by the pore network. Analysis of velocity fluctuations reveals that  rotational and translational kinetic energies are nearly equal for short rods but rotational kinetic energy is relatively suppressed for longer rods.

Because rods can only pass through pores when properly aligned, their orientation dynamics play a central role in determining mobility. Short rods tend to remain vertically aligned, leading to fast, sphere-like percolation. In contrast, long rods sample a wide range of orientations as they navigate the pore structure of the static bed. Consistent with this behavior, the vertical velocity of long rods becomes strongly orientation-dependent.

Trapping is possible once the rod length exceeds a critical value that is about one-half of a sphere diameter for $R=R_t.$ The critical trapping length increases with increasing $R$, which reflects the geometrical conditions required to arrest the rod's motion in the pore structure of the bed. The kinematic transition from percolating to trapping is sharp in terms of vertical displacement with a short stopping distance of $approx 0.5d_L$. Shorter rods are predominantly trapped in nearly vertical configurations through equatorial contacts, whereas longer rods are immobilized over a broader range of orientations. Trapped rods are characterized by a contact number of $Z = 5$, significantly below the isostatic value for spheres and indicating that trapping is governed by geometric confinement and rotational constraints rather than the isostatic limit for frictionless spheres. 

The overall picture that emerges is that rod percolation is governed by a combination of orientation, contact geometry, and pore accessibility. Rods do not behave as spheres: their anisotropy introduces new dynamical constraints, new pathways to arrest motion, and new thresholds for mobility. In particular, elongation induces geometric frustration and enhances multi-point contacts, which contribute to reduced mobility and increased susceptibility to trapping.

These results provide a framework for predicting the transport and segregation of elongated particles in granular materials, with implications for natural systems containing fibers or plant debris and for industrial processes involving pills, grains, or fibrous additives. By revealing how rod length and size ratio combine to produce passing or trapping behavior, this work lays the foundation for extending granular segregation models to a wider range of anisotropic particle shapes. Future work includes incorporating flexible rods, which could bend to navigate pore spaces inaccessible to rigid rods, and adding friction, which introduces tangential resistance and will modify both trapping and orientation dynamics.

\bibliography{Ref}

@article{anzivino2024shear,
  title={Shear flow of non-Brownian rod-sphere mixtures near jamming},
  author={Anzivino, Carmine and Ness, Christopher and Moussa, Amgad Salah and Zaccone, Alessio},
  journal={Phys. Rev. E},
  volume={109},
  number={4},
  pages={L042601},
  year={2024},
  publisher={APS}
}

@article{marschall2018compression,
  title={Compression-driven jamming of athermal frictionless spherocylinders in two dimensions},
  author={Marschall, Theodore and Teitel, Stephen},
  journal={Phys. Rev. E},
  volume={97},
  number={1},
  pages={012905},
  year={2018},
  publisher={APS}
}

@article{wouterse2009contact,
  title={On contact numbers in random rod packings},
  author={Wouterse, Alan and Luding, Stefan and Philipse, Albert P},
  journal={Granul. Matter},
  volume={11},
  number={3},
  pages={169--177},
  year={2009},
  publisher={Springer}
}

@article{blouwolff2006coordination,
  title={The coordination number of granular cylinders},
  author={Blouwolff, J and Fraden, S},
  journal={EPL},
  volume={76},
  number={6},
  pages={1095--1101},
  year={2006}
}

@article{williams2003random,
  title={Random packings of spheres and spherocylinders simulated by mechanical contraction},
  author={Williams, Stephen R and Philipse, Albert P},
  journal={Phys. Rev. E},
  volume={67},
  number={5},
  pages={051301},
  year={2003},
  publisher={APS}
}

@article{petit2022bulk,
  title={Bulk modulus along jamming transition lines of bidisperse granular packings},
  author={Petit, Juan C and Kumar, Nishant and Luding, Stefan and Sperl, Matthias},
  journal={Phys. Rev. E},
  volume={106},
  number={5},
  pages={054903},
  year={2022},
  publisher={APS}
}

@article{ostanin2024rigid,
  title={Rigid clumps in the MercuryDPM particle dynamics code},
  author={Ostanin, Igor and Angelidakis, Vasileios and Plath, Timo and Pourandi, Sahar and Thornton, Anthony and Weinhart, Thomas},
  journal={Comput. Phys. Commun.},
  volume={296},
  pages={109034},
  year={2024},
  publisher={Elsevier}
}

@inproceedings{parteli2013simulation,
  title={DEM simulation of particles of complex shapes using the multisphere method: application for additive manufacturing},
  author={Parteli, Eric JR},
  booktitle={AIP Conference Proceedings},
  volume={1542},
  number={1},
  pages={185--188},
  year={2013},
  organization={American Institute of Physics}
}

@article{vyas2023modelling,
  title={Modelling elastoplastic frictional collisions of ellipsoidal granules with collisional-SPH},
  author={Vyas, Dhairya R and Cummins, Sharen J and Delaney, Gary W and Rudman, Murray and Khakhar, Devang V},
  journal={Adv. Powder Technol.},
  volume={34},
  number={6},
  pages={104028},
  year={2023},
  publisher={Elsevier}
}

@article{kodam2009force,
  title={Force model considerations for glued-sphere discrete element method simulations},
  author={Kodam, Madhusudhan and Bharadwaj, Rahul and Curtis, Jennifer and Hancock, Bruno and Wassgren, Carl},
  journal={Chem. Eng. Sci.},
  volume={64},
  number={15},
  pages={3466--3475},
  year={2009},
  publisher={Elsevier}
}

@article{glielmo2014coefficient,
  title={Coefficient of restitution of aspherical particles},
  author={Glielmo, Aldo and Gunkelmann, Nina and P{\"o}schel, Thorsten},
  journal={Phys. Rev. E},
  volume={90},
  number={5},
  pages={052204},
  year={2014},
  publisher={APS}
}

@article{santos2024protocol,
  title={Protocol-dependent frictional granular jamming simulations: cyclical, compression, and expansion},
  author={Santos, AP and Srivastava, Ishan and Silbert, Leonardo E and Lechman, Jeremy B and Grest, Gary S},
  journal={Front. Soft Matter. },
  volume={3},
  pages={1326756},
  year={2024},
  publisher={Frontiers Media SA}
}

@article{o2003jamming,
  title={Jamming at zero temperature and zero applied stress: The epitome of disorder},
  author={O’hern, Corey S and Silbert, Leonardo E and Liu, Andrea J and Nagel, Sidney R},
  journal={Phys. Rev. E},
  volume={68},
  number={1},
  pages={011306},
  year={2003},
  publisher={APS}
}

@article{luding2022understanding,
  title={Understanding slow compression and decompression of frictionless soft granular matter by network analysis},
  author={Luding, Stefan and Taghizadeh, Kianoosh and Cheng, Chao and Kondic, Lou},
  journal={Soft Matter},
  volume={18},
  number={9},
  pages={1868--1884},
  year={2022},
  publisher={Royal Society of Chemistry}
}

@article{ottino2000mixing,
  title={Mixing and segregation of granular materials},
  author={Ottino, Julio M and Khakhar, Devang V},
  journal={Annu. rev. Fluid Mech.},
  volume={32},
  number={1},
  pages={55--91},
  year={2000},
  publisher={Annual Reviews 4139 El Camino Way, PO Box 10139, Palo Alto, CA 94303-0139, USA}
}

@article{umbanhowar2019modeling,
  title={Modeling segregation in granular flows},
  author={Umbanhowar, Paul B and Lueptow, Richard M and Ottino, Julio M},
  journal={Annu. Rev. Chem. Biomol. Eng.},
  volume={10},
  number={1},
  pages={129--153},
  year={2019},
  publisher={Annual Reviews}
}

@article{jones2021Predicting,
  title = {Predicting segregation of nonspherical particles},
  author = {Jones, Ryan P. and Ottino, Julio M. and Umbanhowar, Paul B. and Lueptow, Richard M.},
  journal = {Phys. Rev. Fluids},
  volume = {6},
  issue = {5},
  pages = {054301},
  numpages = {22},
  year = {2021},
  month = {May},
  publisher = {American Physical Society},
  doi = {10.1103/PhysRevFluids.6.054301},
  url = {https://link.aps.org/doi/10.1103/PhysRevFluids.6.054301}
}

@article{Pohlman2006Surface,
  title = {Surface roughness effects in granular matter: Influence on angle of repose and the absence of segregation},
  author = {Pohlman, Nicholas A. and Severson, Benjamin L. and Ottino, Julio M. and Lueptow, Richard M.},
  journal = {Phys. Rev. E},
  volume = {73},
  issue = {3},
  pages = {031304},
  numpages = {9},
  year = {2006},
  month = {Mar},
  publisher = {American Physical Society},
  doi = {10.1103/PhysRevE.73.031304},
  url = {https://link.aps.org/doi/10.1103/PhysRevE.73.031304}
}

@article{tripathi2013density,
  title={Density difference-driven segregation in a dense granular flow},
  author={Tripathi, Anurag and Khakhar, DV},
  journal={J. Fluid Mech.},
  volume={717},
  pages={643--669},
  year={2013},
  publisher={Cambridge University Press}
}

@article{kudrolli2004size,
  title={Size separation in vibrated granular matter},
  author={Kudrolli, Arshad},
  journal={Rep. Prog. Phys.},
  volume={67},
  number={3},
  pages={209},
  year={2004},
  publisher={IOP Publishing}
}

@article{jaeger1996Granular,
  title = {Granular solids, liquids, and gases},
  author = {Jaeger, Heinrich M. and Nagel, Sidney R. and Behringer, Robert P.},
  journal = {Rev. Mod. Phys.},
  volume = {68},
  issue = {4},
  pages = {1259--1273},
  numpages = {0},
  year = {1996},
  month = {Oct},
  publisher = {American Physical Society},
  doi = {10.1103/RevModPhys.68.1259},
  url = {https://link.aps.org/doi/10.1103/RevModPhys.68.1259}
}

@article{johnson2012grain,
  title={Grain-size segregation and levee formation in geophysical mass flows},
  author={Johnson, CG and Kokelaar, BP and Iverson, Richard M and Logan, M and LaHusen, RG and Gray, JMNT},
  journal={J. Geophys. Res. Earth Surf.},
  volume={117},
  number={F1},
  year={2012},
  publisher={Wiley Online Library}
}

@article{marks2015mixture,
  title={A mixture of crushing and segregation: the complexity of grainsize in natural granular flows},
  author={Marks, Benjy and Einav, Itai},
  journal={Geophys. Res. Lett.},
  volume={42},
  number={2},
  pages={274--281},
  year={2015},
  publisher={Wiley Online Library}
}

@article{jakubowska2021blend,
  title={Blend segregation in tablets manufacturing and its effect on drug content uniformity—a review},
  author={Jakubowska, Emilia and Ciepluch, Natalia},
  journal={Pharmaceutics},
  volume={13},
  number={11},
  pages={1909},
  year={2021},
  publisher={MDPI}
}

@article{dave2015excipient,
  title={Excipient variability and its impact on dosage form functionality},
  author={Dave, Vivek S and Saoji, Suprit D and Raut, Nishikant A and Haware, Rahul V},
  journal={J. Pharm. Sci.},
  volume={104},
  number={3},
  pages={906--915},
  year={2015},
  publisher={Elsevier}
}

@article{jian2019segregation,
  title={Segregation in stored grain bulks: Kinematics, dynamics, mechanisms, and minimization--A review},
  author={Jian, Fuji and Narendran, Ramasamy B and Jayas, Digvir S},
  journal={J. Stored Prod. Res.},
  volume={81},
  pages={11--21},
  year={2019},
  publisher={Elsevier}
}

@article{bemrose1987review,
  title={A review of attrition and attrition test methods},
  author={Bemrose, CR and Bridgwater, J},
  journal={Powder Technol.},
  volume={49},
  number={2},
  pages={97--126},
  year={1987},
  publisher={Elsevier}
}

@book{schulze2008powders,
  title={Powders and bulk solids: behavior, characterization, storage and flow},
  author={Schulze, Dietmar},
  year={2008},
  publisher={Springer}
}

@article{vyas2024impacts,
  title={Impacts of packed bed polydispersity and deformation on fine particle transport},
  author={Vyas, Dhairya R and Gao, Song and Umbanhowar, Paul B and Ottino, Julio M and Lueptow, Richard M},
  journal={AIChE Journal},
  volume={70},
  number={9},
  pages={e18499},
  year={2024},
  publisher={Wiley Online Library}
}

@article{gao2024vertical,
  title={Vertical velocity of a small sphere in a sheared granular bed},
  author={Gao, Song and Ottino, Julio M and Lueptow, Richard M and Umbanhowar, Paul B},
  journal={Phys. Rev. Res.},
  volume={6},
  number={2},
  pages={L022015},
  year={2024},
  publisher={APS}
}

@article{gao2023percolation,
  title={Percolation of a fine particle in static granular beds},
  author={Gao, Song and Ottino, Julio M and Umbanhowar, Paul B and Lueptow, Richard M},
  journal={Phys. Rev. E},
  volume={107},
  number={1},
  pages={014903},
  year={2023},
  publisher={APS}
}

@article{vyas2026fine,
  title={Fine particle percolation dynamics in porous media},
  author={Vyas, Dhairya R and Lueptow, Richard M and Ottino, Julio M and Umbanhowar, Paul B},
  journal={Phys. Rev. Research},
  volume={8},
  number={1},
  pages={013201},
  year={2026},
  publisher={APS}
}

@article{moukarzel1998isostatic,
  title={Isostatic phase transition and instability in stiff granular materials},
  author={Moukarzel, Cristian F},
  journal={Phys. Rev. Lett.},
  volume={81},
  number={8},
  pages={1634},
  year={1998},
  publisher={APS}
}

@article{roux2000geometric,
  title={Geometric origin of mechanical properties of granular materials},
  author={Roux, Jean-No{\"e}l},
  journal={Phys. Rev. E},
  volume={61},
  number={6},
  pages={6802},
  year={2000},
  publisher={APS}
}

@book{duran2012sands,
  title={Sands, powders, and grains: an introduction to the physics of granular materials},
  author={Duran, Jacques},
  year={2012},
  publisher={Springer Science \& Business Media}
}

@article{phillips2006enhanced,
  title={Enhanced mobility of granular mixtures of fine and coarse particles},
  author={Phillips, Jeremy C and Hogg, Andrew J and Kerswell, Richard R and Thomas, Neale H},
  journal={Earth Planet. Sci. Lett.},
  volume={246},
  number={3-4},
  pages={466--480},
  year={2006},
  publisher={Elsevier}
}

@article{dodds1980porosity,
  title={The porosity and contact points in multicomponent random sphere packings calculated by a simple statistical geometric model},
  author={Dodds, JA},
  journal={J. Colloid Interface Sci.},
  volume={77},
  number={2},
  pages={317--327},
  year={1980},
  publisher={Elsevier}
}

@article{lomine2010transit,
  title={Transit time during the interparticle percolation process},
  author={Lomin{\'e}, Franck and Oger, Luc},
  journal={Phys. Rev. E},
  volume={82},
  number={4},
  pages={041301},
  year={2010},
  publisher={APS}
}

@article{rahman2008simulation,
  title={DEM simulation of particle percolation in a packed bed},
  author={Rahman, Mahbubur and Zhu, Haiping and Yu, Aibing and Bridgwater, John},
  journal={Particuology},
  volume={6},
  number={6},
  pages={475--482},
  year={2008},
  publisher={Elsevier}
}

@article{sarkar2017role,
  title={On the role of forces governing particulate interactions in pharmaceutical systems: A review},
  author={Sarkar, Saurabh and Mukherjee, Raj and Chaudhuri, Bodhisattwa},
  journal={Int. J. Pharm.},
  volume={526},
  number={1-2},
  pages={516--537},
  year={2017},
  publisher={Elsevier}
}

@article{flore2009aspects,
  title={Aspects of granulation in the chemical industry},
  author={Flore, Karin and Schoenherr, Michael and Feise, Herrmann},
  journal={Powder Technol.},
  volume={189},
  number={2},
  pages={327--331},
  year={2009},
  publisher={Elsevier}
}

@article{weinhart2020fast,
title = {Fast, flexible particle simulations — An introduction to MercuryDPM},
journal = {Comput. Phys. Commun.},
volume = {249},
pages = {107129},
year = {2020},
issn = {0010-4655},
doi = {https://doi.org/10.1016/j.cpc.2019.107129},
url = {https://www.sciencedirect.com/science/article/pii/S0010465519304357},
author = {Thomas Weinhart and Luca Orefice and Mitchel Post and Marnix P. {van Schrojenstein Lantman} and Irana F.C. Denissen and Deepak R. Tunuguntla and J.M.F. Tsang and Hongyang Cheng and Mohamad Yousef Shaheen and Hao Shi and Paolo Rapino and Elena Grannonio and Nunzio Losacco and Joao Barbosa and Lu Jing and Juan E. {Alvarez Naranjo} and Sudeshna Roy and Wouter K. {den Otter} and Anthony R. Thornton}
}

@article{cundall1979discrete,
	title={A discrete numerical model for granular assemblies},
	author={Cundall, Peter A and Strack, Otto DL},
	journal={G\'eotechnique},
	volume={29},
	number={1},
	pages={47--65},
	year={1979}
}

@article{luding2008cohesive,
  title={Cohesive, frictional powders: contact models for tension},
  author={Luding, Stefan},
  journal={Granul. Matter},
  volume={10},
  number={4},
  pages={235},
  year={2008},
  publisher={Springer}
}

@article{petit2025pressure,
  title={Pressure model and scaling laws in jammed bidisperse granular packings: JC Petit, M. Sperl},
  author={Petit, Juan C and Sperl, Matthias},
  journal={Granul. Matter},
  volume={27},
  number={1},
  pages={23},
  year={2025},
  publisher={Springer}
}

@article{petit2023structural,
  title={Structural transitions in jammed asymmetric bidisperse granular packings: JC Petit and M. Sperl},
  author={Petit, Juan C and Sperl, Matthias},
  journal={Granul. Matter},
  volume={25},
  number={3},
  pages={43},
  year={2023},
  publisher={Springer}
}

@article{petit2020additional,
  title={Additional transition line in jammed asymmetric bidisperse granular packings},
  author={Petit, Juan C and Kumar, Nishant and Luding, Stefan and Sperl, Matthias},
  journal={Phys. Rev. Lett.},
  volume={125},
  number={21},
  pages={215501},
  year={2020},
  publisher={APS}
}
\bibliographystyle{apsrev4-2}

\end{document}